\documentclass[aps,pra,eqsecnum,showpacs,superscriptaddress]{revtex4}
\usepackage{graphicx,color}
\usepackage{amsmath}
\usepackage{amssymb}
\usepackage{bm}

\renewcommand{\d}{\text{d}}
\newcommand{\e}{\text{e}}
\renewcommand{\i}{\text{i}}

\newcommand{\Tr}{\mathop{\text{Tr}}\nolimits}
\newcommand{\ket}[1]{|{#1}\rangle}
\newcommand{\bra}[1]{\langle{#1}|}

\newcommand{\Det}{\mathop{\text{Det}}\nolimits}

\definecolor{dgreen}{rgb}{0,0.5,0}
\newcommand{\dgreen}{\color{dgreen}}
\newcommand{\red}{\color{red}}

\newcommand{\blue}{\color{blue}}
\newcommand{\magenta}{\color{magenta}}
\newcommand{\cyan}{\color{cyan}}

\newcommand{\orange}{\color[named]{Orange}}

\definecolor{delete}{cmyk}{0.5,0,0,0}

\begin{document}

\title{Lateral Effects in Fermion Antibunching}



\author{K. Yuasa}
\altaffiliation[Present Address: ]{Waseda Institute for Advanced Study and Department of Physics, Waseda University, Tokyo 169-8555, Japan}
\affiliation{Dipartimento di Fisica, Universit\`a di Bari, I-70126 Bari, Italy}
\affiliation{Istituto Nazionale di Fisica Nucleare, Sezione di Bari, I-70126 Bari, Italy}

\author{P. Facchi}
\affiliation{Dipartimento di Matematica, Universit\`a di Bari, I-70125 Bari, Italy}
\affiliation{Istituto Nazionale di Fisica Nucleare, Sezione di Bari, I-70126 Bari, Italy}

\author{H. Nakazato}
\affiliation{Department of Physics, Waseda University, Tokyo
169-8555, Japan}

\author{I. Ohba}
\affiliation{Department of Physics, Waseda University, Tokyo
169-8555, Japan}

\author{S. Pascazio}
\affiliation{Dipartimento di Fisica, Universit\`a di Bari, I-70126 Bari, Italy}
\affiliation{Istituto Nazionale di Fisica Nucleare, Sezione di Bari, I-70126 Bari, Italy}

\author{S. Tasaki}
\affiliation{Department of Applied Physics and Advanced Institute
for Complex Systems, Waseda University, Tokyo 169-8555, Japan}



\date[]{March 26, 2008}

\begin{abstract}
Lateral effects are analyzed in the antibunching of a beam of free
non-interacting fermions.
The emission of particles from a source is dynamically described
in a 3D full quantum field-theoretical framework.
The size of the source and the detectors, as well as the temperature
of the source are taken into account and the behavior of the visibility
is scrutinized as a function of these parameters.
\end{abstract}


\maketitle

\section{Introduction}
The wave function of two fermions is antisymmetric under exchange of
the two particles, as a consequence of the Pauli exclusion
principle. For this reason, the probability amplitude for their being
spatially close together is small and their correlated detections
are reduced when compared to a random sequence of classical
particles. This very distinctive quantum feature is named
antibunching and has no classical analog. Notice that in
general the two particles can be emitted from totally incoherent
sources.

The analogous phenomenon for bosons is a cornerstone in the study of
quantum correlations and was first observed in astronomy, where it
is known as the Hanbury Brown-Twiss effect \cite{HBT}. Photon
second-order coherence effects \cite{sudarshan,glauber,bargmann},
yielding bunching, are discussed in physics textbooks
\cite{Loudon,MandelWolf}, and led to novel interesting
applications in quantum imaging \cite{como,UMBC} and lithography
\cite{litho}.

The most relevant difference between the Bose-Einstein and
Fermi-Dirac statistics are the phase space densities (occupation
numbers), that change by several orders of magnitude. In a laser
beam, one obtains values of order $10^{14}$, while typical densities
for thermal light, synchrotron radiation and electrons are of order
$10^{-3}$; finally, for the most advanced neutron sources, one gets
$10^{-15}$. These figures make it very difficult to observe fermion
antibunching. In addition, for charged particles (electrons and
pions), additional Coulomb repulsion effects should be considered,
that tend to reduce the visibility and mask the observation of the
phenomenon.

Quantum correlations have been detected in a series of interesting
experiments: in condensed-matter physics, where the electronic
states are confined within the Fermi surface
\cite{electron1,electron2,electron3}, for
superconductor emitters \cite{Oshima}, in the coincidence spectrum
of neutrons from compound-nuclear reactions at small relative
momentum \cite{tre,qua}, as well as in pion pairs emitted from a
quark-gluon plasma \cite{CERN}. Recently, antibunching was observed
on a beam of thermal neutrons emitted from a nuclear reactor
\cite{IOSFP}. This can be considered as a direct experimental
evidence of free fermion antibunching, in which an ensemble of free
Fermi particles displays quantum coherence effects. Other remarkable
antibunching experiments have been recently reported for neutral
atoms, both in a degenerate atomic Fermi gas \cite{atom} and in
Fermi/Bose gases \cite{He}.

Huge numerical differences in phase space densities, like the
afore-mentioned ones, call for close scrutiny of the theoretical
premises as well as dedicated experimental efforts. Notice that
these quantum statistical effects appear to play a prominent role in
phenomena that are characterized by figures that differ by almost 30
(!) orders of magnitude. The present study is motivated by this
observation. We intend to analyze the antibunching phenomenon in the
correlated detections of two neutral fermions, such as neutrons,
emitted by a generic thermal source at a given temperature. Notice
that bunching effects from (pseudo-)thermal sources still raise
controversial interpretations \cite{thermalcontrov} and are
therefore worth investigating from first principles.

Our main objective will be to analyze the spatial coherence and in
particular the coherence area and volume. Lateral effects are
becoming a critical issue, in view of a new generation of
experiments. They were carefully analyzed in a series of
experimental articles on X-ray bunching \cite{SPring8}.
For the sake of concreteness, we will focus on fermions, but our analysis can be very easily
extended to bosons (by replacing the energy distributions of the source and changing relevant signs in the formulas).

We will treat both the source (a thermal oven) and the particle beam
as fully (second) quantized systems and will study the emission
process at thermal equilibrium, when the beam has reached its
stationary configuration. This approach will have the advantage of
treating both the oven and the fermion beam on an equal footing and
of introducing the properties of the source in a natural way.

\section{Setup and Outline}
\label{sec-outline}
\begin{figure}[b]
\includegraphics[width=0.5\textwidth]{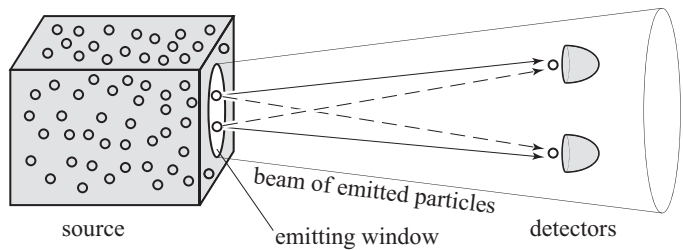}
\caption{Coincidence
between two detectors in the beam of emitted particles: the
interference of the two alternatives yields antibunching.}
\label{fig:setup}
\end{figure}

Before starting a detailed analysis, let us outline the main
features of the setup we have in mind and stress the main points of
our argument. Our setup is the simple one schematically shown in
Fig.\
\ref{fig:setup}. Particles are emitted from a source through a
small window, go through a monochromator (not shown), and are
detected by two detectors. We count the number of coincident
detections. At the initial time $t=0$, the source is in the thermal
equilibrium state at a finite temperature and outside there is the
vacuum. Starting from this initial condition, we shall solve the
dynamics of the emission, so that a stationary beam of particles
will be prepared at $t\to\infty$, after a transient period. The beam
profile will not be added ``by hand," but will be obtained by
solving the equations of motion, so that the coherence properties of
the emitted particles will reflect the dynamics of the emission.

The lateral features of the system affect the antibunching, even
when both detectors are placed on the longitudinal axis and we shall
look at the correlation in the longitudinal direction. To this end,
the lateral size of the detector mouth must be duly taken into
account. We shall therefore implement the lateral resolution of the
detector, as well as the longitudinal one, in the two-particle
distribution function. The variables and parameters that
characterize the setup are summarized in Table
\ref{tab_corr}\@.

We shall start by writing down the Hamiltonian of this many-body
system in Sec.\ \ref{sec:Hamiltonian}\@. This is the crucial part of
the present analysis, since it fully relies on dynamical
consideration. In order to facilitate the introduction of the
characteristics of the source, like temperature, size of the window,
and so on, in a natural way, we shall adopt a two-field approach:
one field describes the particles in the source and the other one
the emitted particles outside. The emission Hamiltonian $\lambda
H_\text{emission}$ (which converts a particle in the source into a
particle outside and \textit{vice versa}) is at the heart of our
analysis and must fully take into account all important features of
the experimental setup, as well as the main characteristics of the
physics of the emission process. The Hamiltonian below will enable
us to discuss the lateral coherence features of the emitted beam,
yet it will be simple enough to be (almost) solvable. As we will
see, the diffraction of the particles emitted through the window
governs the lateral coherence and is controlled by the lateral size
of the emitting window. Once the Hamiltonian is written down, one
has ``only" to solve the equations of motion (and has no ``freedom"
anymore).

The article is organized as follows. The dynamics of the emission is
perturbatively solved in Sec.\
\ref{sec-dynamica}, under the assumption of
weak emissivity, namely weak coupling $\lambda\ll1$, and the
stationary limit $t\to\infty$ realizes a nonequilibrium steady
state. The beam profile thus prepared is studied at a large distance
from the source in Sec.\ \ref{sec-beamprof}\@. We then compute the
two-particle distribution function, or in other words, the
second-order correlation function, defined in Sec.\
\ref{sec-antib}\@. The interplay between the singlet and triplet
contributions determines to which extent the coincidence counts are
reduced (antibunching) when the two detectors are close to each
other. Indeed, the singlet contribution yields bunching and the
triplet one antibunching, with the latter three times larger than
the former. The detector sizes (resolutions) $a$ and $d$ are
implemented into the correlation functions, the saddle-point
approximation is carefully worked out for the case in which the
detectors are placed on the longitudinal $z$ axis, and we obtain a
formula for the normalized two-particle distribution function. The
noncollinear case, with the two detectors placed off the
longitudinal axis, is also discussed.

The antibunching is then discussed and the coherence properties are
clarified in Sec.\ \ref{sec:tempeff}, on the basis of our formula
for the collinear case, and the effect of the temperature of the
source is scrutinized. The temperature effect is shown to be very
weak. The dependence of the antibunching correlation function on the
distance between the two detectors is found to be controlled by the
lateral monochromator window and the longitudinal detector
resolution, while the magnitude of the antibunching effect is
determined by the lateral size of the source. Finally, a variety of
experiments are analyzed in Sec.\ \ref{sec:Exps}, in the light of
the lateral coherence, and the main results are summarized in Sec.\
\ref{sec-concl}\@.
\begin{table}
\caption{Summary of the variables and parameters used in the calculation.}
\label{tab_corr}
\begin{tabular}{cl}
\hline
\hline
  $z$ & longitudinal direction   \\
  $x, y$ & transverse directions    \\
  $w$ & lateral size of the circular emitting window of the source    \\
  $w_z$ &  depth of the emitting region\\
  $\bm{k}_0$  & average momentum at monochromator      \\
  $\delta k_i$ ($i=\perp,z$) &  monochromator window      \\
  $a$ &  lateral size of the circular mouth of the detector  \\
  $d$ & detector resolution in the longitudinal direction      \\
  $0$ (origin) & center of the emitting region      \\
  $\bar{\bm{r}}_i$ ($i=1,2$) & centers of detector apertures      \\
  $\beta$ & inverse temperature of the source      \\
  $\mu$  & Fermi level (in the source)      \\
  $g(\bm{r})$ & emitting window function     \\
  $f(\bm{k})$ & monochromator momentum-window function     \\
  $R_{\bar{\bm{r}}}(\bm{r})$ & detector resolution function    \\ \hline\hline
\end{tabular}
\end{table}

\section{Hamiltonian and State of the Source}
\label{sec-setup}
\label{sec:Hamiltonian}
Let us start with the Hamiltonian: we take
\begin{subequations}
\label{eqn:Hamiltonian}
\begin{gather}
H=H_0+\lambda H_\text{emission},\quad
H_0=H_\text{beam}+H_\text{source},
\intertext{where}
H_\text{beam} =\sum_{\sigma=\uparrow,\downarrow}\int\d^3\bm{k}\,\varepsilon_k
c_{\bm{k}\sigma}^\dag c_{\bm{k}\sigma},\quad H_\text{source}
=\sum_{\sigma=\uparrow,\downarrow}\int\d^3\bm{k}\,\omega_k
a_{\bm{k}\sigma}^\dag a_{\bm{k}\sigma},
\displaybreak[0]\\
H_\text{emission} =\sum_{\sigma=\uparrow,\downarrow}\int\d^3\bm{k}\int\d^3\bm{k}'\,\Bigl(
T_{\bm{k}\bm{k}'}c_{\bm{k}\sigma}^\dag a_{\bm{k}'\sigma}
+T_{\bm{k}\bm{k}'}^*a_{\bm{k}'\sigma}^\dag c_{\bm{k}\sigma}
\Bigr).
\label{eqn:EmissionHamiltonian}
\end{gather}
\end{subequations}
Not only the emitted particles but also the source is treated as a
second quantized dynamical system. The Hamiltonian of the particles
in the source is $H_\text{source}$ and that of the emitted particles
is $H_\text{beam}$. The emission is dynamically described by an
emission Hamiltonian $\lambda H_\text{emission}$: a particle of momentum $\bm{k}'$ and spin $\sigma$ is annihilated
by $a_{\bm{k}'\sigma}$ in the source and is created by
$c_{\bm{k}\sigma}^\dag$ outside with an amplitude $\lambda
T_{\bm{k}\bm{k}'}$, the meaning of which will be described below.
The creation and annihilation operators obey the canonical
anticommutation relations for fermions
\begin{equation}
\{c_{\bm{k}\sigma},c_{\bm{k}'\sigma'}^\dag\}
=\{a_{\bm{k}\sigma},a_{\bm{k}'\sigma'}^\dag\}
=\delta_{\sigma\sigma'}\delta^3(\bm{k}-\bm{k}'),\qquad
\{c_{\bm{k}\sigma},a_{\bm{k}'\sigma'}^\dag\}=0.
\end{equation}
It is assumed that no spin flip occurs during the emission process
and that the emission is irrespective of the spin state of the
particle ($T_{\bm{k}\bm{k'}}$ does not depend on $\sigma$): generalizations to more general cases are straightforward.
The field operator in the configuration space for the emitted
particles is denoted by
\begin{equation}
\psi_{\sigma}(\bm{r},t)
=\int\frac{\d^3\bm{k}}{\sqrt{(2\pi)^3}}\,c_{\bm{k}\sigma}
\e^{\i(\bm{k}\cdot\bm{r}-\varepsilon_kt)}.
\end{equation}
In the following discussion, the dispersion relations are assumed to
be $\varepsilon_k=\omega_k=\bm{k}^2/2m$.

In Eq.\ (\ref{eqn:Hamiltonian}), $\lambda$ is a small parameter, that
will enable us to work in the weak-coupling limit. Although this
approach is familiar in variety of theoretical approaches aimed at
explaining diverse experimental situations, a few words of explanation are
necessary in this case. We have in mind a situation in which an oven
emits a beam of particles through a small aperture (which we refer to
as ``source''). Usually, those particles that leave the source are
monochromatized and can travel in waveguides, undergoing all kinds
of losses. The parameter $\lambda$ globally accounts for all these
diverse processes and $\lambda H_\text{emission}$ simply enables us
to take a particle with approximately the right characteristics in
the oven and put it in the final section of the beam. The smallness of
the opening and the total ``efficiency'' of the emission process
(from the oven to the region of space where the experiment is
practically done, passing through monochromators, optical elements
and/or waveguides and undergoing losses) calls for an approach in
which $\lambda$ is a small parameter. We anticipate that in all
final formulas, where \textit{normalized} distribution functions
will be studied, $\lambda$ will always simplify, making the final
results independent of the details of the apparatus (such as the
monochromatization procedure, reflection and transmission processes,
losses in optical elements and waveguides and so on). Of course, one
must be able to retain all essential elements in the analysis and
final formulas.
The quantity $T_{\bm{k}\bm{k}'}$
in Eq.\ (\ref{eqn:EmissionHamiltonian}) takes into account the
action of the monochromator and the size of the source, and will be
defined in Eq.\ (\ref{eqn:EmissionMatrix}).

The Hamiltonian discussed in this section is to be considered as a
phenomenological transfer Hamiltonian, conveniently tailored in
order to discuss lateral size effects.
It is similar to a ``tunneling'' Hamiltonian (for a two-field formulation of a tunneling
process, see \cite{tunneling}) and can describe the particle emission from a small opening.

\subsection{Emission}
\label{sec:Gaussian}
We consider the following emission process. Only the particles
around the window of the source are emitted outside. That is, a
particle in the momentum state $\ket{\bm{k}}$ (with $k_z>0$) is
annihilated by $a_{\bm{k}\sigma}$ around the window of the source
and is converted into a particle outside by
$c_{\bm{k}'\sigma}^\dag$. The emitting region is specified by a
function $g(\bm{r})$ centered around the window of the source that
characterizes the lateral size of the window. One may further put a
monochromator $f(\bm{k})$ after the emission. The emission Hamiltonian
is then given by
\begin{equation}
H_\text{emission} =\sum_{\sigma=\uparrow,\downarrow}\int\d^3\bm{k}'\int\d^3\bm{k}\,
c_{\bm{k}'\sigma}^\dag\bra{\bm{k}'}f(\bm{k})g(\bm{r})\theta(k_z)\ket{\bm{k}}a_{\bm{k}\sigma}+\text{h.c.}
\end{equation}
with the emission matrix
\begin{equation}
T_{\bm{k}'\bm{k}}
=\bra{\bm{k}'}f(\bm{k})g(\bm{r})\theta(k_z)\ket{\bm{k}}
=f(\bm{k}')\theta(k_z)\int\frac{\d^3\bm{r}}{(2\pi)^3}\,
g(\bm{r})\e^{-\i(\bm{k}'-\bm{k})\cdot\bm{r}}
=f(\bm{k}')\tilde{g}(\bm{k}'-\bm{k})\theta(k_z).
\label{eqn:EmissionMatrix}
\end{equation}
The theta function, $\theta(k_z)=1$ for $k_z>0$ and $\theta(k_z)=0$ for $k_z<0$, accounts for the positivity of the longitudinal momentum $k_z$.
In the following calculation, we assume Gaussian shapes for the emitting
region and the monochromator,
\begin{gather}
g(\bm{r}) =\frac{1}{\sqrt{(2\pi)^3\det\mathcal{W}^2}}
\e^{-\bm{r}\cdot\mathcal{W}^{-2}\bm{r}/2},\quad
\mathcal{W}^2
=\begin{pmatrix}
w^2&0&0\\
0&w^2&0\\
0&0&w_z^2
\end{pmatrix},
\label{eqn:EmittingRegion}\\
f(\bm{k})=\frac{1}{\sqrt[4]{(2\pi)^3\det(\delta\mathcal{K})^2}}
\e^{-(\bm{k}-\bm{k}_0)\cdot(\delta\mathcal{K})^{-2}(\bm{k}-\bm{k}_0)/4},\quad
(\delta\mathcal{K})^2
=\begin{pmatrix}
(\delta k_\perp)^2&0&0\\
0&(\delta k_\perp)^2&0\\
0&0&(\delta k_z)^2
\end{pmatrix},
\label{eqn:Gaussian}
\end{gather}
where $w$ represents the lateral size of the window of
the source, $w_z$ is the depth of the emitting region, and $\delta
k_i$ ($i=\perp,z$) characterize the monochromator.
In the following, we shall take $\bm{k}_0=(0,0,k_0)$.

It is interesting to notice that the non-factorized form
(\ref{eqn:EmissionMatrix}) of the interaction Hamiltonian will
produce the required diffraction effect. (A factorized emission
Hamiltonian would correspond to a point source, irrespectively of
the state before the emission, and would not yield the desired
lateral effect. The choice of the Hamiltonian and the validity of
our working assumptions will be continuously checked throughout the
whole calculation.) A particle with momentum $\bm{k}$ in the source
is converted into a particle with momentum $\bm{k}'$ outside. The
momentum transfer is governed by the Fourier transform
$\tilde{g}(\bm{k}'-\bm{k})$ of the ``interface"  function
$g(\bm{r})$ and is ruled by the size of the window. A smaller window
yields larger momentum transfer and results in a larger divergence
of the emitted beam. This point will be crucial in the following
discussion on the lateral coherence of the beam. The beam profile is
therefore a direct consequence of the dynamics and is not
artificially imposed at the outset. We also notice that the
longitudinal component of momentum is not necessarily preserved
during the emission process, as conservation of the longitudinal
momentum prevents beam divergence. This motivates the choice of the
form factor.

\subsection{State of the Source}
Having written the Hamiltonian of the emission process, it is
straightforward to introduce also the properties of the source. The
initial thermal state of the source at a finite temperature
$\beta^{-1}$ is characterized by
\begin{equation}
\langle
a_{\bm{k}\sigma}^\dag a_{\bm{k}'\sigma'}
\rangle
=N(\omega_k)\delta_{\sigma\sigma'}\delta^3(\bm{k}-\bm{k}'),\qquad
N(\omega)=\frac{1}{\e^{\beta(\omega-\mu)}+1}.
\end{equation}
In the present article, we shall focus for concreteness on the Fermi
distribution. However, we can think of a more general distribution
$N(\omega_k)$. In fact, many of the formulas below remain valid as
long as the initial state is stationary with respect to $H_0$,
admits the Wick decomposition, and $N(\omega_k)$ is a slowly varying function
around $k_0$.

\section{Dynamics of Emission}
\label{sec-dynamica}
The Heisenberg equations of motion read
\begin{equation}
\begin{cases}
\medskip
\displaystyle \i\frac{\d}{\d t}c_{\bm{k}\sigma}(t)
=\varepsilon_kc_{\bm{k}\sigma}(t)
+\lambda\int\d^3\bm{k}'\,T_{\bm{k}\bm{k}'}a_{\bm{k}'\sigma}(t),\\
\displaystyle \i\frac{\d}{\d t}a_{\bm{k}\sigma}(t)
=\omega_ka_{\bm{k}\sigma}(t)
+\lambda\int\d^3\bm{k}'\,T_{\bm{k}'\bm{k}}^*c_{\bm{k}'\sigma}(t).
\end{cases}
\label{eqn:HeisenbergEq}
\end{equation}
By formally integrating the second equation
and inserting it into the first, we get the equation for
$c_{\bm{k}\sigma}(t)$,
\begin{equation}
\i\frac{\d}{\d t}c_{\bm{k}\sigma}(t)
=\varepsilon_kc_{\bm{k}\sigma}(t)
+\lambda\int\d^3\bm{k}'\,T_{\bm{k}\bm{k}'}
\e^{-\i\omega_{k'}t}a_{\bm{k}'\sigma}
-\i\lambda^2\int_0^t\d t'\int\d^3\bm{k}'\,K_{\bm{k}\bm{k}'}(t-t')
c_{\bm{k}'\sigma}(t'),
\label{eqn:NonMarkovEq}
\end{equation}
where
\begin{equation}
K_{\bm{k}\bm{k}'}(t) =\int\d^3\bm{k}''\,T_{\bm{k}\bm{k}''}
\e^{-\i\omega_{k''}t}T_{\bm{k}'\bm{k}''}^*.
\end{equation}
The integro-differential equation (\ref{eqn:NonMarkovEq}) is conveniently solved by Laplace transformation and the solution is given by
\begin{equation}
c_{\bm{k}\sigma}(t)
=\int\d^3\bm{k}'\,G_{\bm{k}\bm{k}'}(t)c_{\bm{k}'\sigma}
-\i\lambda\int_0^t\d t'\int\d^3\bm{k}'\int\d^3\bm{k}''\,
G_{\bm{k}\bm{k}'}(t-t')T_{\bm{k}'\bm{k}''}\e^{-\i\omega_{k''}t'}a_{\bm{k}''\sigma},
\label{eqn:C}
\end{equation}
where
\begin{subequations}
\label{eqn:G}
\begin{gather}
G_{\bm{k}\bm{k}'}(t) =\int_{\text{C}_\text{B}}\frac{\d
s}{2\pi\i}\,\hat{G}_{\bm{k}\bm{k}'}(s)\e^{st},\\
\hat{G}_{\bm{k}\bm{k}'}^{-1}(s)
=(s+\i\varepsilon_k)\delta^3(\bm{k}-\bm{k}')
+\lambda^2\hat{K}_{\bm{k}\bm{k}'}(s),\qquad
\hat{K}_{\bm{k}\bm{k}'}(s)
=\int\d^3\bm{k}''\,\frac{T_{\bm{k}\bm{k}''}T_{\bm{k}'\bm{k}''}^*}{s+\i\omega_{k''}},
 \label{eqn:InverseG}
\end{gather}
\end{subequations}
with $\text{C}_\text{B}$ running parallel to the $s$-imaginary axis (Bromwich path).

\subsection{Nonequilibrium Steady State}
To take the stationary limit $t\to\infty$, it is convenient to move to the interaction picture $\tilde{c}_{\bm{k}\sigma}(t)=\e^{-\i H_0t}c_{\bm{k}\sigma}(t)\e^{\i H_0t}$.
This transformation does not affect the correlation functions of our
initial state (thermal equilibrium inside the source and vacuum
outside), since it is invariant under the free evolution
$\e^{-\i H_0 t}$.
We get
\begin{align}
\langle
c_{\bm{k}_1\sigma_1}^\dag(t) c_{\bm{k}_2\sigma_2}(t)
\rangle
&=\langle \tilde{c}_{\bm{k}_1\sigma_1}^\dag(t)
\tilde{c}_{\bm{k}_2\sigma_2}(t) \rangle\nonumber\\
&\xrightarrow{t\to\infty}
\lambda^2\delta_{\sigma_1\sigma_2}
\int\d^3\bm{k}\,N(\omega_k)
\int_0^\infty\d t_1\,
[G(t_1)T]_{\bm{k}_1\bm{k}}^*\e^{-\i\omega_kt_1}
\int_0^\infty\d t_2\,
[G(t_2)T]_{\bm{k}_2\bm{k}}\e^{\i\omega_kt_2}.
\end{align}
For small $\lambda$ and arbitrary $t$, one obtains (see Appendix
\ref{app:ST})
\begin{equation}
G_{\bm{k}\bm{k}'}(t)=\delta^3(\bm{k}-\bm{k}')\e^{-\i\varepsilon_kt}+O(\lambda^2)
\label{eqn:Glowest0}
\end{equation}
and in the weak-coupling regime we have
\begin{equation}
\langle c_{\bm{k}_1\sigma_1}^\dag(t)
c_{\bm{k}_2\sigma_2}(t) \rangle
\xrightarrow{t\to\infty}
\lambda^2\delta_{\sigma_1\sigma_2}\int\d^3\bm{k}\,
N(\omega_k)
\frac{T_{\bm{k}_1\bm{k}}^*}{\varepsilon_{k_1}-\omega_k+\i0^+}
\frac{T_{\bm{k}_2\bm{k}}}{\varepsilon_{k_2}-\omega_k-\i0^+}
+O(\lambda^4).
\label{eqn:TwoPoint}
\end{equation}
All the other correlation functions are constructed from this
two-point function, through the Wick theorem for an initial thermal
state.

It is instructive to write the correlation function
(\ref{eqn:TwoPoint}) in the configuration space:
\begin{equation}
\langle
\psi_{\sigma_1}^\dag(\bm{r}_1,t)
\psi_{\sigma_2}(\bm{r}_2,t)
\rangle
=\delta_{\sigma_1\sigma_2}
\rho_t^{(1)}(\bm{r}_1|\bm{r}_2)
\xrightarrow{t\to\infty}
\lambda^2\delta_{\sigma_1\sigma_2}
\int\d^3\bm{k}\,N(\omega_k)
\hat{\varphi}_{\bm{k}}^*(\bm{r}_1)
\hat{\varphi}_{\bm{k}}(\bm{r}_2) +O(\lambda^4),
\label{eqn:OneDistr}
\end{equation}
where
\begin{equation}
\hat{\varphi}_{\bm{k}}(\bm{r})
=\int\frac{\d^3\bm{k}'}{\sqrt{(2\pi)^3}\,\i}\,
\frac{T_{\bm{k}'\bm{k}}}{\varepsilon_{k'}-\omega_k-\i0^+}
\e^{\i\bm{k}'\cdot\bm{r}}
\label{eqn:WaveFunction}
\end{equation}
is the Laplace transform of the free evolution of a wave packet
\begin{equation}
\hat{\varphi}_{\bm{k}}(\bm{r},s)
=\int_0^\infty\d t\,\varphi_{\bm{k}}(\bm{r},t)
\e^{-st},\qquad
\varphi_{\bm{k}}(\bm{r},t)
=\int\frac{\d^3\bm{k}'}{\sqrt{(2\pi)^3}}\, T_{\bm{k}'\bm{k}}
\e^{\i(\bm{k}'\cdot\bm{r}-\varepsilon_{k'}t)}
\end{equation}
evaluated on the energy shell $s=-\i\omega_k+0^+$, i.e.\
$\hat{\varphi}_{\bm{k}}(\bm{r})=\hat{\varphi}_{\bm{k}}(\bm{r},-\i\omega_k+0^+)$.
A particle with momentum $\bm{k}$ in the source is diffracted and
propagates outside in the form of the wave packet
$\varphi_{\bm{k}}(\bm{r},t)$. The sum over $\bm{k}$ in formula
(\ref{eqn:OneDistr}) yields the incoherent sum of such wave packets
and a sort of ``density matrix.''

\section{Beam Profile}
\label{sec-beamprof}
It is interesting to observe that the one-particle wave function
(\ref{eqn:WaveFunction}) can be expressed as a superposition of
spherical wave originating from different points of the
emitting region:
\begin{equation}
\hat{\varphi}_{\bm{k}}(\bm{r})
=\theta(k_z)f(-\i\nabla)
\int\frac{\d^3\bm{r}_0}{(2\pi)^3}\,g(\bm{r}_0)
\hat{\varphi}_{\bm{k},\bm{r}_0}^{(0)}(\bm{r}),
\end{equation}
with
\begin{equation}
\hat{\varphi}_{\bm{k},\bm{r}_0}^{(0)}(\bm{r})
=\int\frac{\d^3\bm{k}'}{\sqrt{(2\pi)^3}\,\i}
\frac{1}{\varepsilon_{k'}-\omega_k-\i0^+}
\e^{\i\bm{k}'\cdot(\bm{r}-\bm{r}_0)}
\e^{\i\bm{k}\cdot\bm{r}_0}
=m\sqrt{2\pi}\,\frac{\e^{\i\bm{k}\cdot\bm{r}_0+\i
k|\bm{r}-\bm{r}_0|}}{\i|\bm{r}-\bm{r}_0|}.
\end{equation}
We intend to derive an expression which is valid far from the emitting region.
Equation (\ref{eqn:WaveFunction}) reads
\begin{equation}
\hat{\varphi}_{\bm{k}}(\bm{r})
=\frac{1}{\sqrt{(2\pi)^3}\,\i}
\int_0^\infty\d p\,p^2\int\d^2\hat{\bm{p}}\,
\frac{T_{(p\hat{\bm{p}})\bm{k}}}{\varepsilon_p-\omega_k-\i0^+}
\e^{\i pr(\hat{\bm{p}}\cdot\hat{\bm{r}})}.
\label{4.3}
\end{equation}
For $r\to\infty$, the phase $\hat{\bm{p}}\cdot\hat{\bm{r}}=\cos\theta$ is stationary at $\theta=0$ and $\pi$, and the saddle-point approximation around these points yields
\begin{align}
\hat{\varphi}_{\bm{k}}(\bm{r})
&\sim-\frac{1}{\sqrt{2\pi}}
\int_0^\infty\d p\,p^2\left(
\frac{T_{(p\hat{\bm{r}})\bm{k}}}{\varepsilon_p-\omega_k-\i0^+}
\e^{\i pr}
\int_0^\infty\d u\,u\,
\e^{-pru^2/2}
\right.
\nonumber\\
&\qquad\qquad\qquad\qquad\qquad
\left.
{}-\frac{T_{(-p\hat{\bm{r}})\bm{k}}}{\varepsilon_p-\omega_k-\i0^+}
\e^{-\i pr}
\int_0^\infty\d v\,v\,
\e^{-prv^2/2}
\right)
\nonumber\\
&=-\frac{1}{\sqrt{2\pi}\,r}
\int_{-\infty}^\infty\d p\,p
\frac{T_{(p\hat{\bm{r}})\bm{k}}}{\varepsilon_p-\omega_k-\i0^+}
\e^{\i pr},
\intertext{which asymptotically behaves as}
&{}\sim
m\sqrt{2\pi}\,\theta(k_z)
f(k\hat{\bm{r}})\tilde{g}(\Delta\bm{k}_{\hat{\bm{r}}})
\frac{\e^{\i kr}}{\i r},
\end{align}
where (\ref{eqn:EmissionMatrix}) is substituted for $T_{\bm{p}\bm{k}}$ and
\begin{equation}
\Delta\bm{k}_{\hat{\bm{r}}}=k\hat{\bm{r}}-\bm{k}
\end{equation}
represents the momentum transfer from $\bm{k}$ (before emission)
to that directed towards position $\bm{r}$ with the same magnitude
$k$ (after emission). The Gaussian function
\begin{equation}
\tilde{g}(\Delta\bm{k}_{\hat{\bm{r}}})
=\frac{1}{(2\pi)^3}\e^{-\Delta\bm{k}_{\hat{\bm{r}}}\cdot\mathcal{W}^2\Delta\bm{k}_{\hat{\bm{r}}}/2}
\end{equation}
shows that particles with momentum
$\bm{k}$ in the source prefer to propagate in the same direction
as $\bm{k}$ outside, but with some diffraction determined by the
size of the window of the source.

\section{Correlation Functions}
\label{sec-antib}
We compute the spin-summed one- and two-particle distributions in
the emitted beam, defined respectively by
\begin{equation}
\rho_t^{(1)}(\bm{r})
=\sum_{\sigma=\uparrow,\downarrow}
\langle\psi_\sigma^\dag(\bm{r},t)\psi_\sigma(\bm{r},t)\rangle
=2\rho_t^{(1)}(\bm{r}|\bm{r})
\label{eqn:OneDistriDef}
\end{equation}
and
\begin{align}
\rho_t^{(2)}(\bm{r}_1,\bm{r}_2)
&=\sum_{\sigma_1,\sigma_2=\uparrow,\downarrow}
\langle\psi_{\sigma_1}^\dag(\bm{r}_1,t)\psi_{\sigma_2}^\dag(\bm{r}_2,t)
\psi_{\sigma_2}(\bm{r}_2,t)\psi_{\sigma_1}(\bm{r}_1,t)\rangle
\nonumber\\
&=4\rho_t^{(1)}(\bm{r}_1|\bm{r}_1)\rho_t^{(1)}(\bm{r}_2|\bm{r}_2)
-2\rho_t^{(1)}(\bm{r}_1|\bm{r}_2)\rho_t^{(1)}(\bm{r}_2|\bm{r}_1),
\label{eqn:TwoDistriDef}
\end{align}
where $\rho_t^{(1)}(\bm{r}_1|\bm{r}_2)$ was introduced in (\ref{eqn:OneDistr}).
We are interested in the normalized two-particle distribution function with detector resolutions,
\begin{equation}
\bar{C}_t(\bar{\bm{r}}_1,\bar{\bm{r}}_2)
=\frac{
\bar{\rho}_t^{(2)}(\bar{\bm{r}}_1,\bar{\bm{r}}_2)
}{
\bar{\rho}_t^{(1)}(\bar{\bm{r}}_1)
\bar{\rho}_t^{(1)}(\bar{\bm{r}}_2)
}
=1-\frac{\bar{\mathcal{I}}_t(\bar{\bm{r}}_1,\bar{\bm{r}}_2)}{\bar{\rho}_t^{(1)}(\bar{\bm{r}}_1)\bar{\rho}_t^{(1)}(\bar{\bm{r}}_2)},
\label{eqn:CorrInterfere}
\end{equation}
where
\begin{subequations}
\label{eqn:CorrFuncReso}
\begin{gather}
\bar{\rho}_t^{(1)}(\bar{\bm{r}})
=\int\d^3\bm{r}\,R_{\bar{\bm{r}}}(\bm{r})
\rho_t^{(1)}(\bm{r})
=2\int\d^3\bm{r}\,R_{\bar{\bm{r}}}(\bm{r})
\rho_t^{(1)}(\bm{r}|\bm{r}),\\
\bar{\rho}_t^{(2)}(\bar{\bm{r}}_1,\bar{\bm{r}}_2)
=\int\d^3\bm{r}_1\,R_{\bar{\bm{r}}_1}(\bm{r}_1)
\int\d^3\bm{r}_2\,R_{\bar{\bm{r}}_2}(\bm{r}_2)
\rho_t^{(2)}(\bm{r}_1,\bm{r}_2),
\intertext{and}
\bar{\mathcal{I}}_t(\bar{\bm{r}}_1,\bar{\bm{r}}_2)
=2\int\d^3\bm{r}_1\,R_{\bar{\bm{r}}_1}(\bm{r}_1)
\int\d^3\bm{r}_2\,R_{\bar{\bm{r}}_2}(\bm{r}_2)
\rho_t^{(1)}(\bm{r}_1|\bm{r}_2)\rho_t^{(1)}(\bm{r}_2|\bm{r}_1)
\end{gather}
\end{subequations}
are defined in terms of the resolution function of the detector
$R_{\bar{\bm{r}}}(\bm{r})$, which is assumed to be Gaussian,
\begin{equation}
R_{\bar{\bm{r}}}(\bm{r})
=\frac{1}{\sqrt{(2\pi)^3\det\mathcal{D}^2}}\e^{-(\bm{r}-\bar{\bm{r}})\cdot\mathcal{D}^{-2}(\bm{r}-\bar{\bm{r}})/2},\qquad
\mathcal{D}^2
=\begin{pmatrix}
a^2&0&0\\
0&a^2&0\\
0&0&d^2
\end{pmatrix}.
\label{eqn:GaussianResolution}
\end{equation}
The quantity $a$ characterizes the lateral size of the circular
mouth of the detector and $d$ the resolution in the longitudinal
direction. The ``interference term''
$\bar{\mathcal{I}}_t(\bar{\bm{r}}_1,\bar{\bm{r}}_2)$ gives rise to a
reduction in the two-particle distribution function, that is,
antibunching.
For bosons, the ``$-$'' sign in Eqs.\ (\ref{eqn:TwoDistriDef}) and (\ref{eqn:CorrInterfere}) would be replaced by a ``$+$'' sign, and the coincidence count would be enhanced, exhibiting bunching.
All the formulas below are easily switched to their bosonic counterparts by flipping the negative contribution of the interference term to a positive one.

\subsection{Singlet and Triplet Contributions}
Before we compute the normalized two-particle distribution function
(\ref{eqn:CorrInterfere}), let us look at the structure of the
two-particle distribution (\ref{eqn:TwoDistriDef}) in the stationary
beam:
\begin{align}
\rho_t^{(2)}(\bm{r}_1,\bm{r}_2)
&
\xrightarrow{t\to\infty}
\lambda^4\int\d^3\bm{k}_1\int\d^3\bm{k}_2\,
N(\omega_{k_1})N(\omega_{k_2})\,\Bigl(
4|\hat{\varphi}_{\bm{k}_1}(\bm{r}_1)|^2
|\hat{\varphi}_{\bm{k}_2}(\bm{r}_2)|^2
-2\hat{\varphi}_{\bm{k}_1}^*(\bm{r}_1)
\hat{\varphi}_{\bm{k}_1}(\bm{r}_2)
\hat{\varphi}_{\bm{k}_2}^*(\bm{r}_2)
\hat{\varphi}_{\bm{k}_2}(\bm{r}_1)
\Bigr)\nonumber\displaybreak[0]\\
&=\lambda^4\int\d^3\bm{k}_1\int\d^3\bm{k}_2\,
N(\omega_{k_1})N(\omega_{k_2})\,\Bigl(
3|\Psi_{\bm{k}_1,\bm{k}_2}^{(-)}(\bm{r}_1,\bm{r}_2)|^2
+|\Psi_{\bm{k}_1,\bm{k}_2}^{(+)}(\bm{r}_1,\bm{r}_2)|^2
\Bigr),
\label{eqn:TwoDistriWaveFunc}
\end{align}
where
\begin{equation}
\Psi_{\bm{k}_1,\bm{k}_2}^{(\pm)}(\bm{r}_1,\bm{r}_2)
=\frac{1}{\sqrt{2}}\begin{vmatrix}
\medskip
\hat{\varphi}_{\bm{k}_1}(\bm{r}_1)&
\hat{\varphi}_{\bm{k}_1}(\bm{r}_2)\\
\hat{\varphi}_{\bm{k}_2}(\bm{r}_1)&
\hat{\varphi}_{\bm{k}_2}(\bm{r}_2)
\end{vmatrix}_\pm
\end{equation}
are the symmetrized/antisymmetrized two-particle wave functions.
Formula (\ref{eqn:TwoDistriWaveFunc}) for the two-particle
distribution shows that 3/4 are contributed by the antisymmetric
wave function while 1/4 by the symmetric one. This is because the
thermal source is a complete mixture of the triplet and singlet spin
states, the former being associated with an antisymmetric wave
function in space, while the latter with the symmetric one, for the
state of the fermions as a whole to be antisymmetric.
A similar consideration applies to bosons, for which the symmetrized and antisymmetrized wave functions should be interchanged.

\subsection{Detector Resolution}
Let us now compute the normalized
two-particle distribution function
$\bar{C}_\text{st}(\bar{\bm{r}}_1,\bar{\bm{r}}_2)$ in
(\ref{eqn:CorrInterfere}) in the stationary beam.
The subscript ``st'' will henceforth denote quantities evaluated in the
stationary limit, e.g.\
$\bar{C}_\text{st}(\bar{\bm{r}}_1,\bar{\bm{r}}_2)=\lim_{t\to\infty}\bar{C}_t(\bar{\bm{r}}_1,\bar{\bm{r}}_2)$.
In the main part of this section, we shall employ an approximation which is
    nonsystematic but can nonetheless capture the essential features of the lateral
    effects of $\bar{C}_\text{st}$.
    Its consistency and validity will be examined in Sec.\ \ref{Consistent}\@.

We need to evaluate the following component of the correlation functions in
(\ref{eqn:CorrFuncReso}): by expanding
$\bm{r}$ around the center of the detector,
$\bm{r}=\bar{\bm{r}}+\delta\bm{r}$,
\begin{equation}
\begin{cases}
\displaystyle
r\simeq\bar{r}+\hat{\bar{\bm{r}}}\cdot\delta\bm{r}\\
\displaystyle
\hat{\bm{r}}
\simeq\hat{\bar{\bm{r}}}
+\frac{1}{\bar{r}}\mathcal{P}_{\hat{\bar{\bm{r}}}}\delta\bm{r}\\
\displaystyle
\Delta\bm{k}_{\hat{\bm{r}}}
\simeq\Delta\bm{k}_{\hat{\bar{\bm{r}}}}
+\frac{k}{\bar{r}}\mathcal{P}_{\hat{\bar{\bm{r}}}}\delta\bm{r}
\end{cases}
\quad\text{for}\quad
\delta r\ll\bar{r},
\label{eq:expand1}
\end{equation}
with $\mathcal{P}_{\hat{\bm{r}}}$ a projection operator which projects a vector $\bm{v}$ onto a perpendicular direction to $\hat{\bm{r}}$ by $\mathcal{P}_{\hat{\bm{r}}}\bm{v}=\bm{v}-\hat{\bm{r}}(\hat{\bm{r}}\cdot\bm{v})$, we get
\begin{align}
\int\d^3\bm{r}\,R_{\bar{\bm{r}}}(\bm{r})
\hat{\varphi}_{\bm{k}_1}^*(\bm{r})
\hat{\varphi}_{\bm{k}_2}(\bm{r})
\simeq{}&\frac{m^2}{(2\pi)^5\bar{r}^2}
\theta(k_{1z})
f(k_1\hat{\bar{\bm{r}}})
\theta(k_{2z})
f(k_2\hat{\bar{\bm{r}}})
\e^{-\Delta\bm{k}_{1\hat{\bar{\bm{r}}}}\cdot\mathcal{W}^2\Delta\bm{k}_{1\hat{\bar{\bm{r}}}}/2}
\e^{-\Delta\bm{k}_{2\hat{\bar{\bm{r}}}}\cdot\mathcal{W}^2\Delta\bm{k}_{2\hat{\bar{\bm{r}}}}/2}
\e^{-\i(k_1-k_2)\bar{r}}
\nonumber\\
&\qquad
{}\times\int\d^3\delta\bm{r}\,
\frac{1}{\sqrt{(2\pi)^3\det\mathcal{D}^2}}
\e^{-\delta\bm{r}\cdot\mathcal{B}_{\bar{\bm{r}}}^{-2}\delta\bm{r}/2}
\e^{-(k_1\Delta\bm{k}_{1\hat{\bar{\bm{r}}}}+k_2\Delta\bm{k}_{2\hat{\bar{\bm{r}}}})\cdot\mathcal{W}^2\mathcal{P}_{\hat{\bar{\bm{r}}}}\delta\bm{r}/\bar{r}}
\e^{-\i(k_1-k_2)\hat{\bar{\bm{r}}}\cdot\delta\bm{r}}
\nonumber\displaybreak[0]\\
={}&\mathcal{A}_{\bm{k}_1\bm{k}_2}(\bar{\bm{r}})
\mathcal{Z}_{k_1k_2}(\bar{\bm{r}})
\frac{1}{\bar{r}^2}\e^{-\i(k_1-k_2)\bar{r}},
\label{eq:rphiphi}
\end{align}
where
\begin{subequations}
\begin{align}
\mathcal{A}_{\bm{k}_1\bm{k}_2}(\bar{\bm{r}})
={}&\frac{m^2}{(2\pi)^5}
\frac{1}{\sqrt{\det[1+(k_1^2+k_2^2)\mathcal{D}^2\mathcal{P}_{\hat{\bar{\bm{r}}}}\mathcal{W}^2\mathcal{P}_{\hat{\bar{\bm{r}}}}/\bar{r}^2]}}
\e^{-\Delta\bm{k}_{1\hat{\bar{\bm{r}}}}\cdot\mathcal{W}^2\Delta\bm{k}_{1\hat{\bar{\bm{r}}}}/2}
\e^{-\Delta\bm{k}_{2\hat{\bar{\bm{r}}}}\cdot\mathcal{W}^2\Delta\bm{k}_{2\hat{\bar{\bm{r}}}}/2}
\nonumber\\
&\qquad\quad
{}\times
\e^{(k_1\Delta\bm{k}_{1\hat{\bar{\bm{r}}}}
+k_2\Delta\bm{k}_{2\hat{\bar{\bm{r}}}})
\cdot\mathcal{W}^2\mathcal{P}_{\hat{\bar{\bm{r}}}}\mathcal{B}_{\bar{\bm{r}}}^2\mathcal{P}_{\hat{\bar{\bm{r}}}}\mathcal{W}^2
(k_1\Delta\bm{k}_{1\hat{\bar{\bm{r}}}}
+k_2\Delta\bm{k}_{2\hat{\bar{\bm{r}}}})/2\bar{r}^2}
\e^{\i(k_1-k_2)\hat{\bar{\bm{r}}}
\cdot\mathcal{B}_{\bar{\bm{r}}}^2\mathcal{P}_{\hat{\bar{\bm{r}}}}\mathcal{W}^2
(k_1\Delta\bm{k}_{1\hat{\bar{\bm{r}}}}
+k_2\Delta\bm{k}_{2\hat{\bar{\bm{r}}}})/\bar{r}},
\end{align}
\begin{equation}
\mathcal{Z}_{k_1k_2}(\bar{\bm{r}})
=\theta(k_{1z}) f(k_1\hat{\bar{\bm{r}}})
\theta(k_{2z})
f(k_2\hat{\bar{\bm{r}}})
\e^{-(k_1-k_2)^2
\hat{\bar{\bm{r}}}\cdot\mathcal{B}_{\bar{\bm{r}}}^2\hat{\bar{\bm{r}}}/2},
\end{equation}
and
\begin{equation}
\mathcal{B}_{\bar{\bm{r}}}^{-2}
=\mathcal{D}^{-2}
+(k_1^2+k_2^2)\mathcal{P}_{\hat{\bar{\bm{r}}}}\mathcal{W}^2\mathcal{P}_{\hat{\bar{\bm{r}}}}/\bar{r}^2.
\end{equation}
\end{subequations}
$\mathcal{Z}_{k_1k_2}(\bar{\bm{r}})$ is responsible for the
longitudinal effects and
$\mathcal{A}_{\bm{k}_1\bm{k}_2}(\bar{\bm{r}})$ for the lateral
effects. The one-particle distribution and the interference term in
the two-particle distribution with detector resolutions are then
given by
\begin{equation}
\bar{\rho}_\text{st}^{(1)}(\bar{\bm{r}})
\simeq2\lambda^2
\frac{1}{\bar{r}^2}
\int\d^3\bm{k}\,
N(\omega_k)
\mathcal{A}_{\bm{k}\bm{k}}(\bar{\bm{r}})
\mathcal{Z}_{kk}(\bar{\bm{r}})
\label{eqn:rho1}
\end{equation}
and
\begin{equation}
\bar{\mathcal{I}}_\text{st}(\bar{\bm{r}}_1,\bar{\bm{r}}_2)
\simeq2\lambda^4
\frac{1}{\bar{r}_1^2\bar{r}_2^2}
\int\d^3\bm{k}_1\int\d^3\bm{k}_2\,
N(\omega_{k_1})N(\omega_{k_2})
\mathcal{A}_{\bm{k}_1\bm{k}_2}(\bar{\bm{r}}_1)
\mathcal{A}_{\bm{k}_1\bm{k}_2}^*(\bar{\bm{r}}_2)
\mathcal{Z}_{k_1k_2}(\bar{\bm{r}}_1)
\mathcal{Z}_{k_1k_2}^*(\bar{\bm{r}}_2)
\e^{-\i(k_1-k_2)(\bar{r}_1-\bar{r}_2)}.
\label{eqn:I}
\end{equation}

\subsection{Single-Particle Distribution}
\label{subsec:SPD}
Let us place our detectors on the longitudinal $z$ axis, $\bar{\bm{r}}=(0,0,\bar{z})$.
In this case,
\begin{equation}
\hat{\bar{\bm{r}}}=
\begin{pmatrix}
0\\0\\1
\end{pmatrix}
,\qquad%
\mathcal{P}_{\hat{\bar{\bm{r}}}}=\begin{pmatrix}1&0&0\\0&1&0\\0&0&0\end{pmatrix},
\qquad%
\mathcal{B}_{\bar{\bm{r}}}^2
=\begin{pmatrix}\frac{\displaystyle a^2}{\displaystyle
1+a^2w^2(k_1^2+k_2^2)/\bar{z}^2}&0&0\\
                0&\frac{\displaystyle a^2}{\displaystyle
1+a^2w^2(k_1^2+k_2^2)/\bar{z}^2}&0\\
                0&0&d^2
\end{pmatrix}
\end{equation}
and therefore
\begin{subequations}
\label{eq:A&Z}
\begin{multline}
\mathcal{A}_{\bm{k}_1\bm{k}_2}(\bar{\bm{r}})
=\frac{m^2}{(2\pi)^5}
\frac{1}{1+a^2w^2(k_1^2+k_2^2)/\bar{z}^2}
\e^{-w^2k_{1\perp}^2/2-w_z^2(k_1-k_{1z})^2/2}
\e^{-w^2k_{2\perp}^2/2-w_z^2(k_2-k_{2z})^2/2}
\\
{}\times
\e^{\frac{a^2w^4/2}{\bar{z}^2+a^2w^2(k_1^2+k_2^2)}
(k_1\bm{k}_{1\perp}+k_2\bm{k}_{2\perp})^2},\label{eq:A}
\end{multline}
\begin{equation}
\mathcal{Z}_{k_1k_2}(\bar{\bm{r}})
=\theta(k_{1z}) f(k_1\hat{\bar{\bm{r}}})
\theta(k_{2z})
f(k_2\hat{\bar{\bm{r}}})
\e^{-(k_1-k_2)^2d^2/2}.
\end{equation}
\end{subequations}

Now the single-particle distribution reads
\begin{align}
\bar{\rho}_\text{st}^{(1)}(\bar{\bm{r}})
&\simeq\lambda^2\frac{2m^2}{(2\pi)^5\bar{z}^2}
\int\d^3\bm{k}\,
\frac{
N(\omega_k)\theta(k_z)f^2(k\hat{\bar{\bm{r}}})
}{1+2a^2w^2k^2/\bar{z}^2}
\e^{-\frac{w^2}{1+2a^2w^2k^2/\bar{z}^2}k_\perp^2}\e^{-w_z^2(k-k_z)^2}
\nonumber\displaybreak[0]\\
&=\lambda^2\frac{2m^2}{(2\pi)^4\bar{z}^2}
\int_0^\infty\d k\,k^2
\frac{N(\omega_k)f^2(k\hat{\bar{\bm{r}}})}{1+2a^2w^2k^2/\bar{z}^2}
\int_0^{\pi/2}\d\theta\sin\theta\,\e^{-\frac{w^2k^2}{1+2a^2w^2k^2/\bar{z}^2}\sin^2\!\theta-w_z^2k^2(1-\cos\theta)^2}.
\label{eq:ktrans}
\end{align}
In order to estimate the integral over $\theta$ by a Gaussian approximation, introduce a new integration variable $\Theta$ by
\begin{equation}
\frac{w^2k^2}{1+2a^2w^2k^2/\bar{z}^2}\sin^2\!\theta+w_z^2k^2(1-\cos\theta)^2
=\left(
\frac{w^2k^2}{1+2a^2w^2k^2/\bar{z}^2}+w_z^2k^2\right)\sin^2\!\Theta.
\end{equation}
The above integral over $\theta$ in (\ref{eq:ktrans}) is reduced to the following form
\begin{subequations}
\label{eq:pq}
\begin{gather}
(p+q)\int_0^{\pi/2}\frac{\d\Theta}{\sqrt{p^2\cos^2\!\Theta+q^2\sin^2\!\Theta}}\,\e^{\ln(\sin\Theta\cos\Theta)-(p+q)\sin^2\!\Theta},
\\
p=\frac{w^2k^2}{1+2a^2w^2k^2/\bar{z}^2},
\qquad
q=w_z^2k^2.
\end{gather}
\end{subequations}
Notice that it is important to exponentiate all factors that change considerably in the integration region for the Gaussian approximation to be well posed.
The exponent is expanded around its stationary point
\begin{equation}
\sin^2\!\Theta_0=\frac{1}{2(p+q)}\left(1+p+q-\sqrt{1+(p+q)^2}\right)
\end{equation}
and is approximated by a quadratic function of the form
\begin{multline}
\ln(\sin\Theta\cos\Theta)-(p+q)\sin^2\!\Theta
\\
\sim
\frac{1}{2}\ln\!\left(\frac{\sqrt{1+(p+q)^2}-1}{2(p+q)^2}\right)
-\frac{1}{2}\left(1+p+q-\sqrt{1+(p+q)^2}\right)
-2\sqrt{1+(p+q)^2}(\Theta-\Theta_0)^2.
\end{multline}
The remaining slowly varying factor is estimated at $\Theta_0$ and we obtain
\begin{equation}
\sqrt{\frac{\sqrt{1+(p+q)^2}-1}{p^2+q^2+(p-q)(\sqrt{1+(p+q)^2}-1)}}
\e^{-\frac{1}{2}\left(1+p+q-\sqrt{1+(p+q)^2}\right)}\int_0^{\pi/2}\d\Theta\,\e^{-2\sqrt{1+(p+q)^2}(\Theta-\Theta_0)^2},
\end{equation}
which, for large $p\gg1$, is well approximated by
\begin{equation}
\frac{1}{2\sqrt{p^2+q/2}}\sqrt{\frac{\pi}{\e}}
\sim\frac{1}{2p}\sqrt{\frac{\pi}{\e}}\quad\text{for}\quad p^2\gg q.
\label{eqn:SuddleResult}
\end{equation}
Thus, the single-particle distribution (\ref{eq:ktrans}) is evaluated as
\begin{align}
\bar{\rho}_\text{st}^{(1)}(\bar{\bm{r}})
&\simeq
\lambda^2\frac{2m^2}{(2\pi)^4w^2\bar{z}^2}
\int_0^\infty\d k\,
N(\omega_k)f^2(k\hat{\bar{\bm{r}}})p
(p+q)\int_0^{\pi/2}\frac{\sin\Theta\cos\Theta\,\d\Theta}{\sqrt{p^2\cos^2\!\Theta+q^2\sin^2\!\Theta}}e^{-(p+q)\sin^2\!\Theta}
\nonumber\\
&\simeq
\lambda^2\frac{m^2}{(2\pi)^4w^2\bar{z}^2}\sqrt{\frac{\pi}{\e}}
\int_0^\infty\d k\,
N(\omega_k)f^2(k\hat{\bar{\bm{r}}}).
\label{eq:rho1st}
\end{align}
In the last line, it has been implicitly (and reasonably) assumed
that the monochromator $f(k\hat{\bar{\bm{r}}})$ extracts, in effect,
only those momenta for which the inequality
\begin{equation}
\frac{w^2k^2}{1+2a^2w^2k^2/\bar{z}^2}\gg\max(1,w_zk)
\label{eqn:Condition}
\end{equation}
holds.

If the beam is well monochromatized around a given momentum $k_0$ and the distribution $N(\omega_k)$ is a slowly varying function there, the one-particle distribution function
(\ref{eq:rho1st}) is further estimated for the Gaussian
monochromator (\ref{eqn:Gaussian}) as
\begin{equation}
\bar{\rho}_\text{st}^{(1)}(\bar{\bm{r}})
\simeq
\lambda^2\frac{m^2}{(2\pi)^4w^2\bar{z}^2}\sqrt{\frac{\pi}{\e}}
N(\omega_{k_0})
\int_{-\infty}^\infty\d k\,
f^2(k\hat{\bar{\bm{r}}})
=\lambda^2\frac{m^2}{(2\pi)^5w^2\bar{z}^2(\delta k_\perp)^2}\sqrt{\frac{\pi}{\e}}
N(\omega_{k_0}).
\label{eqn:rho1Lmono2}
\end{equation}

\subsection{Two-Particle Correlation Function }
When the two detectors are placed on the $z$ axis, i.e.\ $\hat{\bar{\bm{r}}}_1=\hat{\bar{\bm{r}}}_2=(0,0,1)=\hat{\bar{\bm{r}}}$, the interference term (\ref{eqn:I}) reads
\begin{align}
\bar{\mathcal{I}}_\text{st}(\bar{\bm{r}}_1,\bar{\bm{r}}_2)
&
\simeq\lambda^4
\frac{2m^4}{(2\pi)^{10}\bar{z}_1^2\bar{z}_2^2}
\int\d^3\bm{k}_1\int\d^3\bm{k}_2\,
\frac{
N(\omega_{k_1}) N(\omega_{k_2})\theta(k_{1z})\theta(k_{2z})f^2(k_1\hat{\bar{\bm{r}}})f^2(k_2\hat{\bar{\bm{r}}})
}{[1+a^2w^2(k_1^2+k_2^2)/\bar{z}_1^2][1+a^2w^2(k_1^2+k_2^2)/\bar{z}_2^2]}
\nonumber\\
&\hspace*{45mm}
{}\times
\e^{-(k_1-k_2)^2d^2}\e^{-\i(k_1-k_2)(\bar{z}_1-\bar{z}_2)}
\e^{-w^2k_{1\perp}^2-w_z^2(k_1-k_{1z})^2}
\e^{-w^2k_{2\perp}^2-w_z^2(k_2-k_{2z})^2}
\nonumber\\
&\hspace*{45mm}
{}\times
\e^{\frac{1}{2}\left(\frac{a^2w^4}{\bar{z}_1^2+a^2w^2(k_1^2+k_2^2)}
+\frac{a^2w^4}{\bar{z}_2^2+a^2w^2(k_1^2+k_2^2)}
\right)
(k_1\bm{k}_{1\perp}+k_2\bm{k}_{2\perp})^2}
\nonumber\\
&
=\lambda^4
\frac{2m^4}{(2\pi)^{10}\bar{z}_1^2\bar{z}_2^2}
\int\d^3\bm{k}_1\int\d^3\bm{k}_2\,
\frac{
N(\omega_{k_1}) N(\omega_{k_2})\theta(k_{1z})\theta(k_{2z})f^2(k_1\hat{\bar{\bm{r}}})f^2(k_2\hat{\bar{\bm{r}}})
}{[1+a^2w^2(k_1^2+k_2^2)/\bar{z}_1^2][1+a^2w^2(k_1^2+k_2^2)/\bar{z}_2^2]}
\nonumber\\
&\hspace*{35mm}
\times
\e^{-(k_1-k_2)^2d^2}\e^{-\i(k_1-k_2)(\bar{z}_1-\bar{z}_2)}
\e^{-p_1k_{1\perp}^2-q_1(k_1-k_{1z})^2}
\e^{-p_2k_{2\perp}^2-q_2(k_2-k_{2z})^2}
\e^{c\bm{k}_{1\perp}\cdot\bm{k}_{2\perp}},
\label{eq:ktransI}
\end{align}
where
\begin{subequations}
\begin{equation}
p_i=w^2-\frac{1}{2}\left(
\frac{a^2w^4}{\bar{z}_1^2+a^2w^2(k_1^2+k_2^2)}
+\frac{a^2w^4}{\bar{z}_2^2+a^2w^2(k_1^2+k_2^2)}
\right)k_i^2,
\quad
q_i=w_z^2
\quad
(i=1,2)
\end{equation}
and
\begin{equation}
c=\left(
\frac{a^2w^4}{\bar{z}_1^2+a^2w^2(k_1^2+k_2^2)}
+\frac{a^2w^4}{\bar{z}_2^2+a^2w^2(k_1^2+k_2^2)}
\right)k_1k_2.
\end{equation}
\end{subequations}
Integrations over the two azimuthal angles around the longitudinal
axis yield a modified Bessel function of the first kind
\begin{equation}
2\pi\int_0^{2\pi}\d\phi_{12}\,\e^{ck_{1\perp}k_{2\perp}\cos\phi_{12}}
=(2\pi)^2\sum_{n=0}^\infty\frac{1}{(n!)^2}
\left(\frac{c^2}{4}\right)^nk_{1\perp}^{2n}k_{2\perp}^{2n}
=(2\pi)^2I_0(ck_{1\perp}k_{2\perp}).
\end{equation}
The remaining integrations over $k_{1\perp}$ and $k_{2\perp}$ can be performed just like before and we obtain [cf.\ (\ref{eq:ktrans}) and (\ref{eqn:SuddleResult})]
\begin{align}
&\bar{\mathcal{I}}_\text{st}(\bar{\bm{r}}_1,\bar{\bm{r}}_2)
\nonumber\\
&\ \ %
\simeq\lambda^4
\frac{2m^4}{(2\pi)^8\bar{z}_1^2\bar{z}_2^2}
\int_0^\infty\d k_1\,k_1^2\int_0^\infty\d k_2\,k_2^2
\e^{-(k_1-k_2)^2d^2}\e^{-\i(k_1-k_2)(\bar{z}_1-\bar{z}_2)}
\nonumber\\
&\hspace*{50mm}
\times
\sum_{n=0}^\infty\frac{(c^2/4)^n}{(n!)^2}
\prod_{i=1,2}
\frac{
N(\omega_{k_i})f^2(k_i\hat{\bar{\bm{r}}})}{1+a^2w^2(k_1^2+k_2^2)/\bar{z}_i^2}
\nonumber\\
&\hspace*{70mm}
\times
\left(
-\frac{\partial}{\partial p_i}
\right)^n
\int_0^{\pi/2}\d\theta_i\sin\theta_i\,
\e^{-p_ik_i^2\sin^2\!\theta_i-q_ik_i^2(1-\cos\theta_i)^2}
\nonumber\displaybreak[0]\\
&\ \ %
\simeq\lambda^4
\frac{2m^4}{(2\pi)^8\bar{z}_1^2\bar{z}_2^2}
\int_0^\infty\d k_1\,k_1^2\int_0^\infty\d k_2\,k_2^2
\e^{-(k_1-k_2)^2d^2}\e^{-\i(k_1-k_2)(\bar{z}_1-\bar{z}_2)}
\nonumber\\
&\hspace*{50mm}
\times
\sum_{n=0}^\infty\frac{(c^2/4)^n}{(n!)^2}
\prod_{i=1,2}
\frac{
N(\omega_{k_i})f^2(k_i\hat{\bar{\bm{r}}})}{1+a^2w^2(k_1^2+k_2^2)/\bar{z}_i^2}
\left(
-\frac{\partial}{\partial p_i}
\right)^n
\frac{1}{2p_ik_i^2}\sqrt{\frac{\pi}{\e}}
\nonumber\displaybreak[0]\\
&\ \ %
=\lambda^4
\frac{m^4}{2(2\pi)^8\bar{z}_1^2\bar{z}_2^2}
\frac{\pi}{\e}
\int_0^\infty\d k_1\int_0^\infty\d k_2\,
\frac{
N(\omega_{k_1})N(\omega_{k_2})f^2(k_1\hat{\bar{\bm{r}}})f^2(k_2\hat{\bar{\bm{r}}})}{[1+a^2w^2(k_1^2+k_2^2)/\bar{z}_1^2][1+a^2w^2(k_1^2+k_2^2)/\bar{z}_2^2]}
\frac{\e^{-(k_1-k_2)^2d^2}\e^{-\i(k_1-k_2)(\bar{z}_1-\bar{z}_2)}}{p_1p_2-c^2/4}
\nonumber\displaybreak[0]\\
&\ \ %
=\lambda^4
\frac{m^4}{2(2\pi)^8w^4\bar{z}_1^2\bar{z}_2^2}\frac{\pi}{\e}
\int_0^\infty\!\d k_1\int_0^\infty\!\d k_2\,
\frac{
N(\omega_{k_1})N(\omega_{k_2})f^2(k_1\hat{\bar{\bm{r}}})f^2(k_2\hat{\bar{\bm{r}}})
}{1+a^2w^2(k_1^2+k_2^2)(1/2\bar{z}_1^2+1/2\bar{z}_2^2)}
\e^{-(k_1-k_2)^2d^2}\e^{-\i(k_1-k_2)(\bar{z}_1-\bar{z}_2)}.
\label{eq:Iint}
\end{align}
For the well-monochromatized case,
\begin{align}
\bar{\mathcal{I}}_\text{st}(\bar{\bm{r}}_1,\bar{\bm{r}}_2)
\simeq\lambda^4&
\frac{m^4}{2(2\pi)^8w^4\bar{z}_1^2\bar{z}_2^2}
\frac{\pi}{\e}
\frac{N^2(\omega_{k_0})}{1+a^2w^2k_0^2(1/\bar{z}_1^2+1/\bar{z}_2^2)}
\nonumber\\
&{}\times\int_{-\infty}^\infty\d k_1
\int_{-\infty}^\infty\d k_2\,
f^2(k_1\hat{\bar{\bm{r}}}) f^2(k_2\hat{\bar{\bm{r}}})
\e^{-(k_1-k_2)^2d^2}
\e^{-\i(k_1-k_2)(\bar{z}_1-\bar{z}_2)}
\nonumber\displaybreak[0]\\
=\lambda^4&
\frac{m^4}{2(2\pi)^{10}w^4\bar{z}_1^2\bar{z}_2^2(\delta k_\perp)^4}
\frac{\pi}{\e}
\frac{N^2(\omega_{k_0})}{1+a^2w^2k_0^2(1/\bar{z}_1^2+1/\bar{z}_2^2)}
\frac{1}{\sqrt{1+4(\delta k_z)^2d^2}}
\exp\!\left(
-\frac{(\bar{z}_1-\bar{z}_2)^2}{1/(\delta k_z)^2+4d^2}
\right),
\label{eq:Ist}
\end{align}
and we end up with the analytical formula for the normalized
two-particle distribution function:
\begin{equation}
\bar{C}_\text{st}(\bar{\bm{r}}_1,\bar{\bm{r}}_2)
=1-\frac{1}{2}
\frac{1}{1+a^2w^2k_0^2(1/\bar{z}_1^2+1/\bar{z}_2^2)}
\frac{1}{\sqrt{1+4(\delta k_z)^2d^2}}
\exp\!\left(
-\frac{(\bar{z}_1-\bar{z}_2)^2}{1/(\delta k_z)^2+4d^2}
\right).
\label{eqn:Analytical}
\end{equation}
This is our central result.
Its bosonic counterpart is readily obtained by just flipping the ``$-$'' sign in front of the second term.

Before studying this formula numerically, it is worth analyzing its range of validity.

\subsection{Consistency of the Approximations}
\label{Consistent}
One might wonder if the approximation and procedure we adopted when
we performed the integration over $\bm{r}$ in (\ref{eq:rphiphi}) are
self-consistent, because, as some careful reader might have
realized, we have partly kept second-order terms in $\delta\bm{r}$
in the exponent of the integrand of (\ref{eq:rphiphi}), while only
first-order corrections were considered in the expansions
(\ref{eq:expand1}). Actually, we implicitly assumed that
$\delta\bm{r}=\bm{r}-\bar{\bm{r}}$ is a small quantity, in order to keep
only second-order terms in $\delta\bm{r}$ in the exponent, so that
the integrals could be evaluated by Gaussian integrations. Stated
differently, it is not clear whether we are allowed to expand the
exponent of the integrand around $\bar{\bm{r}}$, since this is the
stationary point of the exponent of $R_{\bar{\bm{r}}}(\bm{r})$, but
is not necessarily that of the integrand. In order for the
approximation be consistent, we first have to find the true
stationary or saddle point of the exponent of the integrand,
$\bm{r}_\text{s}$, and then expand it around ${\bm{r}}_\text{s}$,
keeping all second-order terms in $\bm{r}-{\bm{r}}_\text{s}$.

It is not difficult to derive the relation that the saddle point $\bm{r}_\text{s}$ of the exponent of the integrand
\begin{multline}
F(\bm{r})=
-\frac{1}{2}(\bm{r}-\bar{\bm{r}})\cdot\mathcal{D}^{-2}(\bm{r}-\bar{\bm{r}})
-\frac{1}{4}(k_1\hat{\bm{r}}-\bm{k}_0)\cdot(\delta\mathcal{K})^{-2}(k_1\hat{\bm{r}}-\bm{k}_0)
-\frac{1}{4}(k_2\hat{\bm{r}}-\bm{k}_0)\cdot(\delta\mathcal{K})^{-2}(k_2\hat{\bm{r}}-\bm{k}_0)
\\
-\frac{1}{2}\Delta\bm{k}_{1\hat{\bm{r}}}\cdot\mathcal{W}^2\Delta\bm{k}_{1\hat{\bm{r}}}
-\frac{1}{2}\Delta\bm{k}_{2\hat{\bm{r}}}\cdot\mathcal{W}^2\Delta\bm{k}_{2\hat{\bm{r}}}
-\i(k_1-k_2)r
\label{eq:exponentF}
\end{multline}
has to satisfy. This reads
\begin{align}
0=\nabla F(\bm{r})\Bigr|_{\bm{r}_\text{s}}
=-\mathcal{D}^{-2}(\bm{r}_\text{s}-\bar{\bm{r}})
&{}-\frac{k_1}{2r_\text{s}}\mathcal{P}_{\hat{\bm{r}}_\text{s}}(\delta\mathcal{K})^{-2}(k_1\hat{\bm{r}}_\text{s}-\bm{k}_0)
-\frac{k_2}{2r_\text{s}}\mathcal{P}_{\hat{\bm{r}}_\text{s}}(\delta\mathcal{K})^{-2}(k_2\hat{\bm{r}}_\text{s}-\bm{k}_0)
\nonumber\\
&{}-\frac{k_1}{r_\text{s}}\mathcal{P}_{\hat{\bm{r}}_\text{s}}\mathcal{W}^2\Delta\bm{k}_{1\hat{\bm{r}}_\text{s}}
-\frac{k_2}{r_\text{s}}\mathcal{P}_{\hat{\bm{r}}_\text{s}}\mathcal{W}^2\Delta\bm{k}_{2\hat{\bm{r}}_\text{s}}
-\i(k_1-k_2)\hat{\bm{r}}_\text{s}.
\end{align}
The saddle point $\bm{r}_\text{s}$ is therefore the solution of the
equation
\begin{equation}
\bm{r}_\text{s}=\bar{\bm{r}}-\bm{e}(\bm{r}_\text{s}),
\end{equation}
with
\begin{subequations}
\begin{gather}
\bm{e}(\bm{r})=\frac{1}{r}\mathcal{D}^2\mathcal{P}_{\hat{\bm{r}}}\bm{u}(\bm{r})+\i(k_1-k_2)\mathcal{D}^2\hat{\bm{r}},\\
\bm{u}(\bm{r})=(k_1^2+k_2^2)
\left(
\mathcal{W}^2+\frac{1}{2}(\delta\mathcal{K})^{-2}
\right)\hat{\bm{r}}
-\frac{k_1+k_2}{2}(\delta\mathcal{K})^{-2}\bm{k}_0
-\mathcal{W}^2(k_1\bm{k}_1+k_2\bm{k}_2).
\end{gather}
\end{subequations}
This equation is iteratively solved to yield, after the first iteration,
\begin{equation}
\bm{r}_\text{s}
\simeq\bar{\bm{r}}-\bm{e}(\bar{\bm{r}})
=\bar{z}\left[
\left(
1-\i(k_1-k_2)\frac{d^2}{\bar{z}^2}
\right)\hat{\bar{\bm{r}}}
+\frac{a^2w^2}{\bar{z}^2}(k_1\bm{k}_1+k_2\bm{k}_2)_\perp
\right].
\end{equation}
Notice that for this iterative solution to be a good approximation,
its deviation from $\bar{\bm{r}}$ must be a small quantity relative
to $\bar{z}=|\bar{\bm{r}}|$. Actually, this expression is still valid,
after the second iteration, up to the second order in $a/\bar{z}$ and
$d/\bar{z}$.

Since we obtain
\begin{equation}
r_\text{s}
\simeq\bar{z}\left(
1-\i(k_1-k_2)\frac{d^2}{\bar{z}^2}
\right)
\quad\text{and}\quad
\hat{\bm{r}}_\text{s}
\simeq\hat{\bar{\bm{r}}}+\frac{a^2w^2}{\bar{z}^2}(k_1\bm{k}_1+k_2\bm{k}_2)_\perp
\end{equation}
up to $O(a^2/\bar{z}^2)$ and $O(d^2/\bar{z}^2)$, the saddle-point value of the exponent, $F(\bm{r}_\text{s})$, is easily estimated to be
\begin{multline}
F(\bm{r}_\text{s})
={-\frac{1}{2}}(k_1-k_2)^2d^2
-\i(k_1-k_2)\bar{z}
-\frac{1}{2}w^2(k_{1\perp}^2+k_{2\perp}^2)
+\frac{a^2w^4}{2\bar{z}^2}(k_1\bm{k}_1+k_2\bm{k}_2)_\perp^2\\
-\frac{(k_1-k_0)^2}{4(\delta k_z)^2}
-\frac{(k_2-k_0)^2}{4(\delta k_z)^2}
-\frac{1}{2}w_z^2[(k_1-k_{1z})^2+(k_2-k_{2z})^2].
\label{eq:Frs}
\end{multline}
We then approximate the exponent by a quadratic form
\begin{equation}
F(\bm{r})
\simeq F(\bm{r}_\text{s})
+\frac{1}{2!}\sum_{i,j}
(x_i-x_{\text{s}i})(x_j-x_{\text{s}j})
\left.\frac{\partial^2F(\bm{r})}{\partial x_i\,\partial x_j}\right|_{\bm{r}_\text{s}}
\end{equation}
to perform the Gaussian integration.
The covariance matrix (the coefficient matrix of the quadratic term) reads
\begin{align}
\frac{1}{2!}\sum_{i,j}
(x_i-x_{\text{s}i})(x_j-x_{\text{s}j})
\left.\frac{\partial^2F(\bm{r})}{\partial x_i\,\partial x_j}\right|_{\bm{r}_\text{s}}
&
=-\begin{pmatrix}
(\bm{r}-\bm{r}_\text{s})_\perp^\text{t}&
z-z_\text{s}
\end{pmatrix}
\begin{pmatrix}
\mathcal{M}&\bm{b}\\
\bm{b}^\text{t}&\alpha
\end{pmatrix}
\begin{pmatrix}
(\bm{r}-\bm{r}_\text{s})_\perp\\
z-z_\text{s}
\end{pmatrix}\nonumber\displaybreak[0]\\
&
=-\alpha\left(
(z-z_\text{s})+\frac{1}{\alpha}(\bm{r}-\bm{r}_\text{s})_\perp\cdot\bm{b}\right)^2
-(\bm{r}-\bm{r}_\text{s})_\perp\cdot\left(\mathcal{M}-\frac{\bm{b}\bm{b}^\text{t}}{\alpha}\right)
(\bm{r}-\bm{r}_\text{s})_\perp,
\end{align}
which yields, after integrations over $\bm{r}-\bm{r}_\text{s}$,
\begin{equation}
\sqrt{\frac{\pi}{\alpha}}
\sqrt{\frac{\pi^2}{\det(\mathcal{M}-\bm{b}\bm{b}^\text{t}/\alpha)}}.
\end{equation}
Within the validity of the approximation adopted here, this factor is estimated to be
\begin{equation}
\sqrt{(2\pi)^3\det\mathcal{D}^2}
\left(
1+\frac{a^2w^2(k_1^2+k_2^2)}{\bar{z}^2}
\Lambda_{\bm{k}_1\bm{k}_2}
+\frac{\i(k_1-k_2)a^2}{\bar{z}-\i(k_1-k_2)d^2}
\right)^{-1},
\label{eq:detfactor}
\end{equation}
where
\begin{equation}
\Lambda_{\bm{k}_1\bm{k}_2}
=1+\frac{1}{2w^2(\delta k_\perp)^2}
-\frac{1}{2w^2(\delta k_z)^2}\left(
1-\frac{k_0(k_1+k_2)}{k_1^2+k_2^2}
\right)
-\frac{w_z^2}{w^2}\left(
1-\frac{k_1k_{1z}+k_2k_{2z}}{k_1^2+k_2^2}
\right).
\label{eq:h}
\end{equation}

It is worth mentioning that we have obtained the same value as
before [see, for example, (\ref{eq:A&Z})] for the saddle-point value
of the exponent (\ref{eq:Frs}), within the validity of our
approximation [up to $O(a^2/\bar{z}^2)$ and $O(d^2/\bar{z}^2)$]. The
only correction to $\mathcal{A}_{\bm{k}_1\bm{k}_2}(\bm{r})$ in
(\ref{eq:A}) is to replace its denominator by the quantity in the
brackets in (\ref{eq:detfactor})
\begin{equation}
1+\frac{a^2w^2(k_1^2+k_2^2)}{\bar{z}^2}
\ \ \longrightarrow\ \ %
1+\frac{a^2w^2(k_1^2+k_2^2)}{\bar{z}^2}
\left(
\Lambda_{\bm{k}_1\bm{k}_2}
+\frac{2}{w^2(k_1^2+k_2^2)}
\right)
-\frac{3d^2}{\bar{z}^2}
+\frac{\i(k_1-k_2)a^2}{\bar{z}-\i(k_1-k_2)d^2}.
\end{equation}
If we take, however, the same Gaussian approximation for the
momentum integrations as in Sec.\ \ref{subsec:SPD}, by which the
last term in the right hand side of (\ref{eq:h}) is estimated to be
$\sim-w_z^2/2w^4(k_1^2+k_2^2)$, and consider the
well-monochromatized case, few corrections remain. Actually, it can
be easily shown that the single-particle distribution
(\ref{eqn:rho1Lmono2}) has further to be divided by a factor
$1+(a^2/\bar{z}^2)[k_0^2/(\delta k_\perp)^2]$ and the denominator in
the interference term (\ref{eq:Ist}) has to be corrected by
\begin{equation}
1+a^2w^2k_0^2\left(\frac{1}{\bar{z}_1^2}+\frac{1}{\bar{z}_2^2}
\right)\quad
\longrightarrow\quad
1+a^2w^2k_0^2\left(\frac{1}{\bar{z}_1^2}+\frac{1}{\bar{z}_2^2}
\right)\left(
1+\frac{1}{w^2(\delta k_\perp)^2}
\right).
\end{equation}
It is remarkable and important to note that these corrections do not change the final result for the normalized two-particle distribution function (\ref{eqn:Analytical}), illustrating the consistency and validity of the approximation and procedure we adopted there.

\subsection{Lateral Correlation at the Lowest Order for a Noncollinear Arrangement}
\label{sec:first}
So far, we have investigated the lateral effects when the source and
the two detectors are collinearly arranged. In order to bring
antibunching to light, the second-order expansion of the exponent
around the saddle point was necessary. It is interesting to note
that, if the two detectors are placed off the longitudinal $z$ axis,
lateral effects can be seen even in the lowest order with respect to
$1/\bar{r}_1$ and $1/\bar{r}_2$. Indeed, when the detectors are
placed on $
\bar{\bm{r}}_1=(
\bar{r}_1 \sin\Theta_\text{d} \cos\Phi,
\bar{r}_1 \sin\Theta_\text{d} \sin\Phi,
\bar{r}_1 \cos\Theta_\text{d})
$, $
\bar{\bm{r}}_2 = (
-\bar{r}_2 \sin\Theta_\text{d} \cos\Phi,
-\bar{r}_2 \sin\Theta_\text{d} \sin\Phi,
\bar{r}_2 \cos\Theta_\text{d})
$, the normalized two-particle distribution function
(\ref{eqn:CorrInterfere}) is given by
\begin{equation}
\bar{C}_\text{st}(\bar{\bm{r}}_1,\bar{\bm{r}}_2)
=
1-
\frac{D_1(\Theta_\text{d})}{2
\sqrt{
D_2(\Theta_\text{d})
D_3(\Theta_\text{d})
}}
\exp\!
\left(
- \frac{w^2k_0^2(\delta k_\perp)^2\sin^2\!2\Theta_\text{d}}{ 2D_1(\Theta_\text{d})
D_2(\Theta_\text{d})
}
-
\frac{
(\delta k_z)^2(\delta k_\perp)^2
(\bar{r}_1-\bar{r}_2)^2
}{
D_3(\Theta_\text{d})
}
\right),
\label{OffCorrelation}
\end{equation}
provided the beam of particles is well-monochromatized and the two
inequalities $w\gg w_z$, $wk\gg 1$ are satisfied. Here the auxiliary
functions $D_1$, $D_2$, and $D_3$ are defined as
\begin{subequations}
\begin{align}
D_1(\Theta_\text{d})
&=(\delta k_z)^2 \sin^2\!\Theta_\text{d}
+(\delta k_\perp)^2\cos^2\!\Theta_\text{d}
+2w_z^2(\delta k_z)^2(\delta k_\perp)^2
(1-\cos\Theta_\text{d})^2,
\label{D_1}
\\
D_2(\Theta_\text{d})
&=
D_1(\Theta_\text{d}) +2w^2(\delta k_z)^2(\delta k_\perp)^2 \sin^2\! \Theta_\text{d}
,
\label{D_2}
\\
D_3(\Theta_\text{d})
&=
D_2(\Theta_\text{d})+4a^2(\delta k_z)^2(\delta k_\perp)^2 \sin^2\!\Theta_\text{d}
+4d^2(\delta k_z)^2(\delta k_\perp)^2\cos^2\!\Theta_\text{d}
.
\label{D_3}
\end{align}
\end{subequations}
As can clearly be seen, the two-particle distribution does depend on
the lateral size $w$ of the source. The derivation is discussed in
Appendix \ref{AppST}\@.

Now we consider two particular cases. In the first case where $\Theta_\text{d}=0$, one has
$D_1(0)=D_2(0)=(\delta k_\perp)^2$ and $D_3(0)=(\delta k_\perp)^2 [1+4(\delta k_z)^2d^2]$.
Thus, (\ref{OffCorrelation})
reduces to the lowest-order terms of (\ref{eqn:Analytical}) with respect to $1/\bar{z}_1$ and $1/\bar{z}_2$:
\begin{equation}
\bar{C}_\text{st}(\bar{\bm{r}}_1,\bar{\bm{r}}_2)
=1-\frac{1}{2\sqrt{1+4(\delta k_z)^2d^2}}
\exp\!\left(-\frac{(\bar{z}_1-\bar{z}_2)^2}{1/(\delta k_z)^2+4d^2}\right),
\label{OffCorrelation1}
\end{equation}
where we have set $\bar{z}_1=\bar{r}_1$ and $\bar{z}_2=\bar{r}_2$.
In the second case, where $\bm{r}_1=(\bar{x},\bar{y},\bar{z})$ and $\bm{r}_2=(-\bar{x},-\bar{y},\bar{z})$,
the two-particle distribution reads
\begin{equation}
\bar{C}_\text{st}(\bar{\bm{r}}_1,\bar{\bm{r}}_2)
=
1-
\frac{\widetilde{D}_1(\bar{x},\bar{y},\bar{z})}{
2
\sqrt{
\widetilde{D}_2(\bar{x},\bar{y},\bar{z})
\widetilde{D}_3(\bar{x},\bar{y},\bar{z})
}}
\exp\!
\left(
- \frac{2w^2k_0^2(\delta k_\perp)^4(\bar{x}^2+\bar{y}^2)\bar{z}^2}{\widetilde{D}_1(\bar{x},\bar{y},\bar{z})
\widetilde{D}_2(\bar{x},\bar{y},\bar{z})
}
\right),
\label{OffCorrelation2}
\end{equation}
where the functions $\widetilde{D}_1$, $\widetilde{D}_2$, and
$\widetilde{D}_3$ are
\begin{subequations}
\begin{align}
\widetilde{D}_1(\bar{x},\bar{y},\bar{z})
&=
\bar{r}^2 D_1(\Theta_\text{d})
=(\delta k_z)^2 (\bar{x}^2+\bar{y}^2)
+ (\delta k_\perp)^2\bar{z}^2
+2w_z^2(\delta k_z)^2(\delta k_\perp)^2(\bar{r}-\bar{z})^2 ,
\label{2D_1}
\\
\widetilde{D}_2(\bar{x},\bar{y},\bar{z})
&=
\bar{r}^2D_2(\Theta_\text{d})=
\widetilde{D}_1(\bar{x},\bar{y},\bar{z})
+2w^2(\delta k_z)^2(\delta k_\perp)^2 (\bar{x}^2+\bar{y}^2)
,
\label{2D_2}
\\
\widetilde{D}_3(\bar{x},\bar{y},\bar{z})
&=
\bar{r}^2 D_3(\Theta_\text{d})=
\widetilde{D}_2(\bar{x},\bar{y},\bar{z})
+4(\delta k_z)^2(\delta k_\perp)^2[a^2(\bar{x}^2+\bar{y}^2)
+ d^2 \bar{z}^2],
\label{2D_3}
\end{align}
\end{subequations}
and $\bar{r}=\sqrt{\bar{x}^2+\bar{y}^2+\bar{z}^2}$. The geometrical
configuration of noncollinear detectors is not the usual one and
will not be discussed any further. It might be important for
electron antibunching from superconducting emitters \cite{inprep}.

\section{Longitudinal, Lateral, and Temperature Effects on Antibunching}
\label{sec:tempeff}
\begin{figure}
\includegraphics[width=0.4\textwidth]{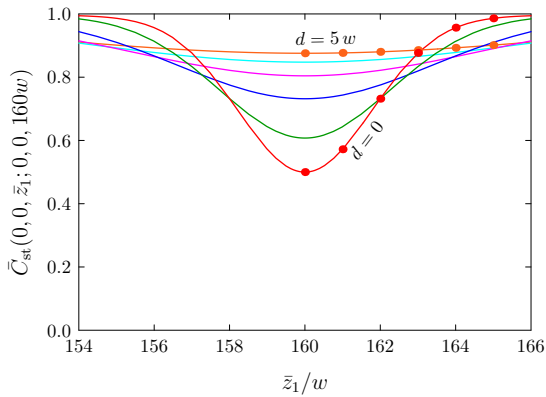}
\caption{(Color online) Normalized two-particle distribution function
$\bar{C}_\text{st}(0,0,\bar{z}_1;0,0,160w)$ vs
the longitudinal detector coordinate
$\bar{z}_1$.
$\bar{C}_\text{st}$ is evaluated
in the stationary state on the
basis of the expressions (\ref{eq:rho1st}) and (\ref{eq:Iint}),
with the Gaussian detectors located at $\bar{\bm{r}}_1=(0,0,\bar{z}_1)$
and $\bar{\bm{r}}_2=(0,0,160w)$.
We focus here on the effect of the longitudinal resolution of the detector, $d$.
The parameters are $k_0=20\,w^{-1}$, $\delta k_z=0.5\,w^{-1}$, $a=0$, and from bottom to top (in the dip)
$d/w={\red0},{\dgreen1},{\blue2},{\magenta3},{\cyan4},{\orange5}$ 
(in units $\hbar=1$ and $m=1$).  We set $\beta=5\,mw^2/\hbar^2$ (a low temperature) and the
Fermi level $\mu=(k_0+\delta k_z)^2/2m\simeq210\,\hbar^2/mw^2$ (just above the momentum window).
Note that in this case $p\simeq400$ in (\ref{eq:pq})
and the condition (\ref{eqn:Condition}) imposes $w_z\ll20\,w$.
The values based on the numerical integrations of (\ref{eq:ktrans})
and (\ref{eq:ktransI}) without the Gaussian approximation are also shown
by dots for 
$d={\red0}$ and ${\orange5}\,w$. 
These were checked to be independent
of $w_z$ for small $w_z$.
}
\label{fig:Longitudinal}
\end{figure}

The normalized two-particle distribution function
$\bar{C}_\text{st}(\bar{\bm{r}}_1,\bar{\bm{r}}_2)$, evaluated on the
basis of the expressions (\ref{eq:rho1st}) and (\ref{eq:Iint}), is
shown in Figs.\ \ref{fig:Longitudinal}--\ref{fig:Temp}. The
two-particle distribution (the number of coincidence counts) is
always suppressed when the two detectors are close together. The
dips in the figures represent antibunching.

It is clear from the expression (\ref{eqn:TwoDistriDef}) for the
two-particle distribution that the number of coincidences is reduced
to one half of that naively expected on the basis of the counts by
the single detectors, when the two (ideal) point detectors are at
the same point. This is understood by the expression
(\ref{eqn:TwoDistriWaveFunc}): the triplet spin states accompany the
antisymmetric wave function in space, yielding antibunching, while
the singlet spin state accompanies the symmetric wave function in
space, yielding bunching. The interplay of these contributions
(three fourths from the antibunching and one fourth from bunching)
results in the minimum value $0.5\,(=1-3/4+1/4)$ of the normalized
two-particle distribution function. The width of the dip, on the
other hand, is governed by the width of the spectrum of the emitted
particles, as is clear from the analytical formula
(\ref{eqn:Analytical}) or from the Fourier-integral representation
of the interference term in (\ref{eqn:I}).

Let us next look at the effects of the detector resolutions. Not
only the longitudinal resolution of the detectors $d$ (Fig.\
\ref{fig:Longitudinal}) but also the lateral size of the detectors
$a$ affects the visibility of the antibunching (Fig.\
\ref{fig:Lateral}), although we are looking at the coincidences
between the two detectors located along the longitudinal axis. The
width of the dip is broadened by the longitudinal resolution of the
detectors $d$, while it is not by the lateral size $a$.

The analytical formula (\ref{eqn:Analytical}) clearly reveals how
the resolutions of the detectors affect the coincidence counts. It
also shows that the detectors can be regarded as point detectors when
\begin{equation}
a\ll\frac{\bar{z}}{\sqrt{2}\,wk_0},\qquad d\ll\frac{1}{2\,\delta
k_z},
\end{equation}
and these quantities characterize the lateral and longitudinal
coherence lengths, respectively.
Clearly, the above conditions agree with those derived in classical textbooks \cite{MandelWolf}.

The temperature of the source affects antibunching in a way that deserves a few words of explanation. The visibility (namely, the depth of the dip) is temperature independent, as a consequence of the antisymmetry of the fermionic state \textit{\`a~la} Pauli's principle: indeed, antisymmetry is \textit{exact}, both for pure and mixed states, and is preserved even at very high temperatures, where the fermionic state is totally mixed. See Fig.\ \ref{fig:Temp}(a). 
This is clear in our formulas and figures: the prefactor $-1/2$ does not depend on any details of the source  and the experimental setup.
See e.g.\ Eq.\ (\ref{eqn:TwoDistriWaveFunc}): in the second line, irrespectively of the temperature distribution $N(\omega_k)$, when $|\bm{r}_1-\bm{r}_2|\to\infty$ the interference term (second term in brackets) goes to zero, while at $\bm{r}_1=\bm{r}_2$ it equals half of the background term (first term in brackets).
On the other hand, the width of the dip depends on temperature and even strongly by the location of the monochromator window with respect to the Fermi level, as can be clearly seen in Fig.\ \ref{fig:Temp}(b). 
As a consequence, the dip becomes narrower as temperature is increased and the effects of antibunching becomes more difficult to detect.

\section{Application to experiments}
\label{sec:Exps}
It is useful to summarize the meaning of our analysis.
Equations (\ref{eqn:OneDistriDef}) and (\ref{eqn:TwoDistriDef})
yield the one- and two-particle distributions in the beam, that are
expressed in terms of the fermionic operators. Equations
(\ref{eqn:OneDistr}) and (\ref{eqn:TwoDistriWaveFunc}) then express
these quantities in terms of the ``wave functions''
$\hat{\varphi}_{\bm{k}}(\bm{r})$ and of the temperature dependent
function $N(\omega_k)$. If the Fermi distribution is plugged in, all
formulas apply to fermions: otherwise the analysis in Secs.\ \ref{sec-dynamica}--\ref{sec:tempeff} is general (modulo some sign changes) and can be applied to
bosons as well.

It is interesting to apply our final result (\ref{eqn:Analytical})
to some interesting experimental situations. It is necessary to
stress that our analysis is strictly valid only for experiments such
that the beam of emitted particles travels in vacuum. If this
situation closely resembles the experimental one, then our equations
apply; otherwise, additional care is required in order to explain
the experimental data. In some experiments, like those in which
correlation in the current intensities are observed
\cite{electron1,electron2}, our formulas cannot be applied and a
different analysis is required \cite{inprep}.

Let us start from an analysis of the electron experiment
\cite{electron3}. One infers the values $a\simeq 2\,\text{mm}$,
$w \simeq 18\,\text{nm}$, $k_0\simeq10^{11}\,\text{m}^{-1}$,
$z\simeq 10\,\text{cm}$. By plugging these values in Eq.\
(\ref{eqn:Analytical}), one sees that the first of the two factors
multiplying the exponential is of order $10^{-4}$.
Moreover, the coherence time is $\Delta t_\text{coh}\simeq 32\,\text{fs}$, while the
response time of the detectors $\Delta t_\text{det}\simeq
26\,\text{ps}$, which yield a value $\simeq 1/300$ for the second
factor in front of the exponential. The global factor multiplying
the exponential is therefore of order $10^{-6}$, which makes the
observation of the phenomenon quite complicated. Indeed, the authors
had to apply a lateral magnification technique (nominally of order
$\gtrsim 10^4$) in order to observe antibunching. Notice
also that in our formulas the Coulomb repulsion is neglected. This
is a delicate issue that would require additional investigation.

Let us now look at the neutron experiment \cite{IOSFP}. The
relevant values are $a\simeq 1\,\text{cm}$, $w \simeq
1\,\text{cm}$ (a mosaic crystal was used in order to reflect the beam into the apparatus), $k_0\simeq10^{10}\,\text{m}^{-1}$,
$z\simeq 10\,\text{m}$. The beam coming out of an oven travels in
waveguides for about $100\,\text{m}$, is then monochromatized through
back scattering by a prefect crystal and illuminates the whole mosaic
crystal on a region of order few $\text{cm}^2$, the back reflection
being coherent only on regions of order $\simeq
\mu\text{m}^2$ that are uniformly distributed in the whole
volume. By plugging the numerical values of the parameters in Eq.\
(\ref{eqn:Analytical}), the first factor is of order $10^{-10}$.
Moreover, by comparing the coherence time of the neutron
wave packet $\Delta t_\text{coh}\simeq 20\,\text{ns}$ with the
response times of the detectors $\Delta t_\text{det}\simeq
0.1\,\mu\text{s}$ (two different types of detectors were used, with
response times that differ by a factor 10--20), one obtains a second
factor of order $\gtrsim 10^{-1}$ (or smaller by a factor 10 for the
other type of detectors), yielding a very small antibunching dip.
It is interesting to observe that if we take $w \simeq
1\,\mu\text{m}$ (the size of a monocrystal in the mosaic), by Eq.\
(\ref{eqn:Analytical}) the first factor is of order $10^{-2}$, the
second factor remains identical and one obtains an antibunching dip of a few percent, which can be brought to light by deconvolution and is in agreement with the experimental data. An
exhaustive analysis of the physical effects of the mosaic crystal
used in back reflection is involved and will be presented elsewhere.
\begin{figure}
\vspace*{8truemm}
\begin{tabular}{r@{\qquad\qquad}r}
(a)\hspace*{4truemm}\mbox{}&(b)\hspace*{4truemm}\mbox{}\\[-8truemm]
\includegraphics[width=0.4\textwidth]{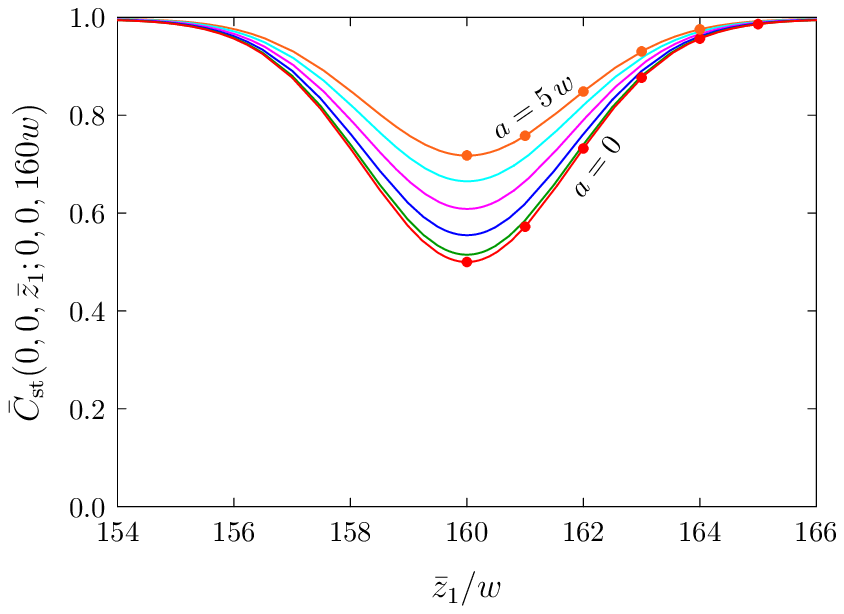}&
\includegraphics[width=0.4\textwidth]{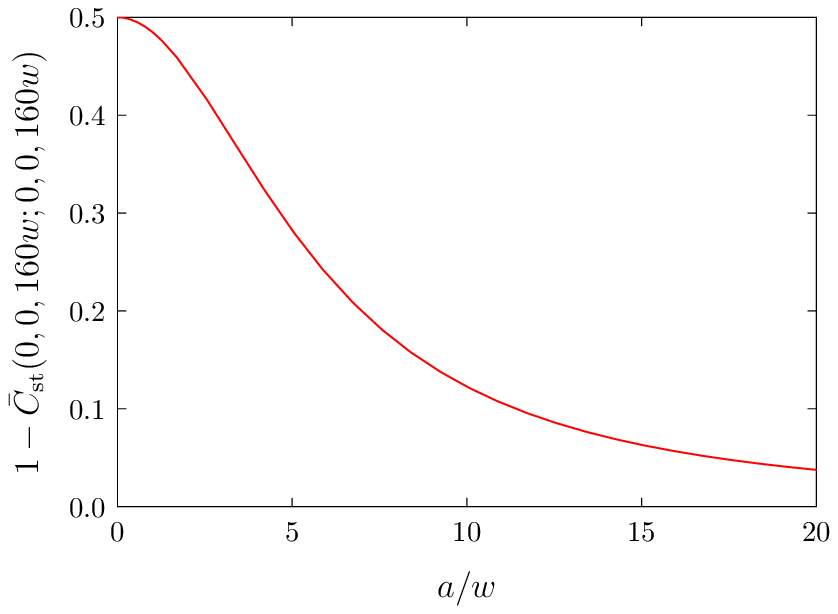}
\end{tabular}
\caption{(Color online) (a) Same as in Fig.\ \ref{fig:Longitudinal} but with $d=0$ and
from bottom to top 
$a/w={\red0},{\dgreen1},{\blue2},{\magenta3},{\cyan4},{\orange5}$.
We study here the effect of the lateral size of the detector, $a$.
Note that $p\simeq225$ for 
$a={\orange5}\,w$ 
in (\ref{eq:pq}) and the condition
(\ref{eqn:Condition}) imposes $w_z\ll11\,w$.
The values based on the numerical integrations of (\ref{eq:ktrans})
and (\ref{eq:ktransI})
without the Gaussian approximation are also shown by dots for 
$a={\red0}$ and ${\orange5}\,w$ 
and were checked to be independent of $w_z$ for small $w_z$.
(b) Depth of the antibunching dip, $1-\bar{C}_\text{st}(0,0,160w;0,0,160w)$,
as a function of $a$.  All the parameters are the same as in (a).}
\label{fig:Lateral}
\end{figure}
\begin{figure}
\vspace*{8truemm}
\begin{tabular}{r@{\qquad\qquad}r}
(a)\hspace*{4truemm}\mbox{}&(b)\hspace*{4truemm}\mbox{}\\[-8truemm]
\includegraphics[width=0.4\textwidth]{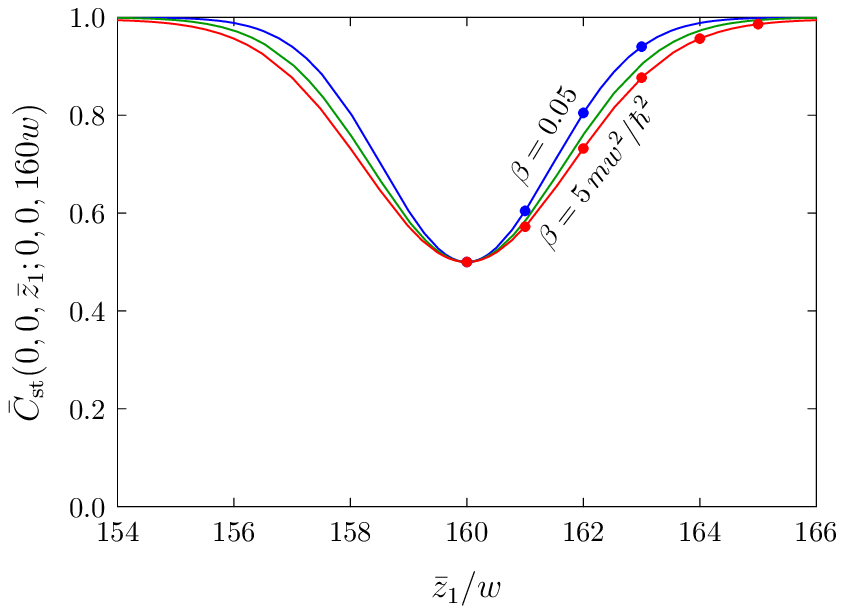}&
\includegraphics[width=0.36\textwidth]{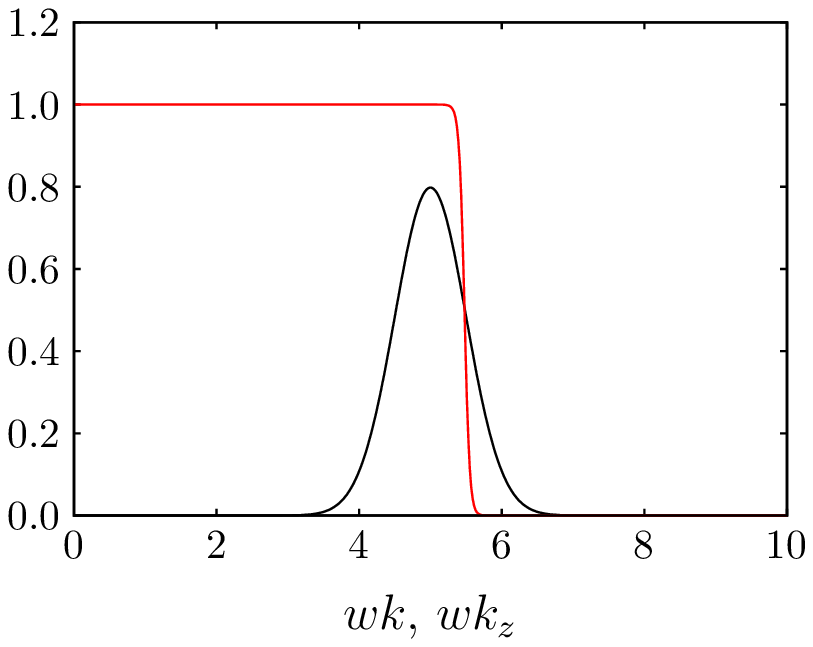}
\end{tabular}
\caption{(Color online) (a) Same as in Fig.\ \ref{fig:Longitudinal} but
with $a=0$, $d=0$, and from bottom to top 
$\beta/(mw^2/\hbar^2)={\red5},{\dgreen0.2},{\blue0.05}$.
We analyze here the effect of the temperature, $\beta^{-1}$.
The values based on the numerical integrations of (\ref{eq:ktrans})
and (\ref{eq:ktransI}) without the Gaussian approximation are also
shown for 
$\beta/(mw^2/\hbar^2)={\red5}$ and ${\blue0.05}$ 
by dots, and
were checked to be independent of $w_z$ for small $w_z$.
(b) Particle emission (from the oven into the beam) is proportional
to the overlap of the form factor of the monochromator (bell-shaped curve)
and the Fermi distribution function in the oven (step-like function).
The effective width of the overlap is inversely proportional to the
width of the (antibunching) dip in (a).
}
\label{fig:Temp}
\end{figure}

Another interesting experiment is that performed with X-rays
\cite{SPring8}. An important part of this experiment is devoted to
the analysis of the lateral coherence features of the beam. The
setup involves the values $k_0\simeq7\cdot10^{10}\,\text{m}^{-1}$,
$z\simeq 70\,\text{m}$, while the detector size was changed in the
range $a\simeq 10$--$400\,\mu\text{m}$. The authors observed a
reduction of the bunching peak very similar to that described in our
Fig.\
\ref{fig:Lateral}(b), as a function of the size of the detector
mouth $a$, and estimated the lateral coherence length of the beam
from the width of the plot, obtaining a source size of order
$w\simeq20\,\mu\text{m}$. The longitudinal coherence length is known
to be $\Delta\ell_\text{coh}\simeq2\,\text{mm}$, while the length of
the photon bunch is $\Delta\ell_\text{det}\simeq5\,\text{mm}$: by
plugging these values in our Eq.\ (\ref{eqn:Analytical}), the first
factor is $\simeq0.9$ for $a\simeq10\,\mu\text{m}$, while the second
one is $\simeq0.4$, so that $(1/2) \cdot 0.4 \cdot 0.9 \simeq 0.18$,
which explain well the observed data (a positive bump $\simeq
0.25$.)

Finally, it is interesting to look at some experiments done with
thermal optical photon, in order to study both their correlations
and imaging. Experiments of this kind were widely debated during the
last few years \cite{como,UMBC,thermalcontrov,sergienko}. We shall
focus on the experiment \cite{examplethermal}, in which the source
is a He-Ne laser (in other similar experiments a green-doubled
Nd:YAG laser was used). One estimates $a\simeq 30\,\mu\text{m}$, $w
\simeq 1\,\text{mm}$, $k_0\simeq10^{7}\,\text{m}^{-1}$, $z\simeq
150\,\text{mm}$. The pseudothermal source is obtained by randomizing
the phase of the photon field by means of a rotating ground-glass
disk (so that the expression random source would probably be more
appropriate). The first factor in Eq.\ (\ref{eqn:Analytical}) is
therefore $\simeq 0.1$ while the second one is essentially 1
($\Delta t_\text{det}\simeq 500\,\text{ps}$, yielding
$\Delta\ell_\text{det}\simeq0.15\,\text{m}$, which is to be compared
to the much larger coherence length of the laser). This yields an
overall dip of good visibility. Unfortunately a quantitative
comparison with the experimental data is difficult because the
authors, being interested in the change of the width of the second
order correlation function with the source size, only plotted a
\emph{(re)normalized} correlation function.

This brief summary of experimental data shows that our analysis and
final formulas agree well with most experiments performed so far,
with different particles. In some cases a comparison is more
complicated and/or requires additional information.
One phenomenon
that we find of interest, but still lacks experimental confirmation,
is our prediction that for fermionic systems we expect that the
visibility of second-order interference effects should show
\textit{no dependence on temperature}, as explained in Fig.\
\ref{fig:Temp}. We emphasized at the end of Sec.\
\ref{sec:tempeff}  that this is due to the exactness of
Pauli's principle (yielding perfect antisymmetrization) even for
mixed states.

\section{Conclusions, Comments, and Perspectives}
\label{sec-concl}
We analyzed antibunching in a beam of non-interacting fermions and
investigated the behavior of the visibility as a function of the
size of the source and the detectors, as well as the temperature of
the source. These parameters are critical and play a prominent role
in experimental applications. Our analysis makes use of Gaussian
functions both for the emitting region of the source and the
detector, and is adapted to an approximately cylindrically symmetric
situation, with circular detector placed close to the longitudinal
axis. This is the relevant situation in most experiments, in
particular with neutrons and electrons (where however additional
Coulomb effects, as well as more specific emission features of the
source need to be scrutinized). It is also worth
noticing that in the observation of pion correlations
\cite{CERN} the experimental data have been exploited to determine the dimension
and the expansion dynamics of the pion ``source" (fireball produced
in central Pb-Pb collisions, which is expected to be a droplet of
quark gluon plasma at the freeze-out point). Clearly, our approach
can prove to be quite useful in such a situation as well, when
source size and temperature are not known.

Let us look at possible applications and future perspectives. First
of all, we emphasize that although our analysis was performed for
fermions, all formulas can be easily translated to the case of
bosons, enabling one to scrutinize interesting novel experiments in
quantum imaging and lithography. Some recent applications make use
of chaotic or pseudo-thermal light sources, to which our formalism
immediately applies \cite{como,UMBC,litho}.

Other possible applications are in solid state physics, where, as a
consequences of the symmetrization procedure of the many-body wave
function, entanglement should be present in bulk matter, raising
delicate problems in relation to its detection and extraction
\cite{vedral,dechiara,Giovannetti2}. Since one should
get entangled neutron pairs within the coherence volume of the wave
packet (antibunching being observed within the same coherence
volume), these pairs could be used as very efficient ``probes" for
entanglement in solids in future experiments. This is clearly
relevant for quantum information processing and for tests of the
Bell inequalities.

Finally, we mention some interesting speculations in neutrino
physics and the structure of the universe, where, as a consequence
of entanglement formation, the hypothesis of a collisionless fluid
of classical point particles can be critically re-examined, yielding
a ``quantum overpressure," with significant consequences during the
non-linear structure formation epoch at low redshifts
\cite{neutrino}. It is remarkable that independent ideas can bear
consequences in very diverse fields.

\acknowledgments
We would like to thank B.\ Cho, M.\ D'Angelo, D. Di Bari, S.\
Filipp, B.\ Ghidini, Y.\ Hasegawa, M.\ Iannuzzi, L.~A.\ Lugiato, C.\
Oshima, H.\ Rauch, F.\ Sacchetti, G. Scarcelli and Y.\ Shih for
discussions and pertinent remarks. This work is supported by the
bilateral Italian Japanese Projects II04C1AF4E on ``Quantum
Information, Computation and Communication" of the Italian Ministry
of Instruction, University and Research, and the Joint Italian Japanese Laboratory on ``Quantum
Information and Computation" of the Italian Ministry for Foreign
Affairs, by the European Union through the Integrated Project
EuroSQIP, by the Grants for The 21st Century COE Program ``Holistic
Research and Education Center for Physics of Self-Organization
Systems" and for the ``Academic Frontier'' Project at Waseda
University and the Grants-in-Aid for the COE Research
``Establishment of Molecular Nano-Engineering by Utilizing
Nanostructure Arrays and Its Development into Micro-Systems" at
Waseda University (No.\ 13CE2003) and for Young Scientists (B) (No.\
18740250) from the Ministry of Education, Culture, Sports, Science
and Technology, Japan, and by the Grants-in-Aid for Scientific
Research (C) (Nos.\ 17540365 and 18540292) from the Japan Society
for the Promotion of Science.

\appendix
\section{$G(t)$ in the Weak-Coupling Regime}
\label{app:ST}
Let us prove (\ref{eqn:Glowest0}). We write the matrix
(\ref{eqn:InverseG}) as
\begin{equation}
\hat{G}^{-1}(s)=\hat{D}(s)+\lambda^2\hat{K}(s),\qquad
\hat{D}_{\bm{k}\bm{k}'}(s)=(s+\i\varepsilon_k)\delta^3(\bm{k}-\bm{k}').
\end{equation}
Let us show that
\begin{equation}
[1+\lambda^2\hat{K}(s)\hat{D}^{-1}(s)]^{-1}
\end{equation}
does not have poles in the weak-coupling regime, i.e.
\begin{equation}
\Det[1+\lambda^2\hat{K}(s)\hat{D}^{-1}(s)]\neq0 ,
\label{eqn:NoPole}
\end{equation}
for a sufficiently small $\lambda$. Indeed, the determinant is
evaluated as
\begin{align}
\Det[1+\lambda^2\hat{K}(s)\hat{D}^{-1}(s)]
&=\e^{\Tr\log[1+\lambda^2\hat{K}(s)\hat{D}^{-1}(s)]},
\intertext{which reads, in the weak-coupling regime,}
&=1+\lambda^2\Tr[\hat{K}(s)\hat{D}^{-1}(s)]+O(\lambda^4)
\nonumber\displaybreak[0]\\
&=1+\lambda^2\int\d^3\bm{k}\,\frac{\hat{K}_{\bm{k}\bm{k}}(s)}{s+\i\varepsilon_k}+O(\lambda^4)\nonumber\\
&=1+\lambda^2\int\d^3\bm{k}\int\d^3\bm{k}'\,
\frac{|T_{\bm{k}\bm{k}'}|^2}{(s+\i\varepsilon_k)(s+\i\omega_{k'})}+O(\lambda^4).
\end{align}
Let us look for a zero of the determinant on the first Riemannian
sheet, by putting $s=-\i\omega-\gamma/2$:
\begin{equation}
\Det[1+\lambda^2(\hat{K}\hat{D}^{-1})(-\i\omega-\gamma/2)]
=1-\lambda^2\int\d^3\bm{k}\int\d^3\bm{k}'\,
\frac{|T_{\bm{k}\bm{k}'}|^2}{(\varepsilon_k-\omega+\i\gamma/2)(\omega_{k'}-\omega+\i\gamma/2)}
+O(\lambda^4). \label{eqn:DetZero}
\end{equation}
Assume now that the square of the emission matrix in the energy
representation,
\begin{equation}
\Gamma(\omega,\omega')=(2\pi)^2\int\d^3\bm{k}\int\d^3\bm{k}'\,
|T_{\bm{k}\bm{k}'}|^2\delta(\varepsilon_k-\omega)\delta(\omega_{k'}-\omega'),
\label{eqn:gammaom}
\end{equation}
does not contain any ``nonlocal'' part $\delta(\omega-\omega')$ and
has the properties
\begin{equation}
\Gamma(\omega,\omega')\to0\quad\text{for}\quad \omega,\omega'\to
0,\infty
\end{equation}
[assuming some good continuity property for
$\Gamma(\omega,\omega')$]. For such a ``reasonable'' emission matrix
$T_{\bm{k}\bm{k}'}$, the integral in the second term of
(\ref{eqn:gammaom})
\begin{equation}
\int_0^\infty\frac{\d\omega_1}{2\pi}\int_0^\infty\frac{\d\omega_2}{2\pi}
\frac{\Gamma(\omega_1,\omega_2)}{(\omega_1-\omega+\i\gamma/2)(\omega_2-\omega+\i\gamma/2)}
\end{equation}
is convergent for any $\omega$ and $\gamma$, and the determinant can
always be made non-zero by choosing a sufficiently small $\lambda$.
The pole of
\begin{equation}
\hat{G}(s)
=\hat{D}^{-1}(s)[1+\lambda^2\hat{K}(s)\hat{D}^{-1}(s)]^{-1},
\end{equation}
therefore, comes from the first factor
$\hat{D}_{\bm{k}\bm{k}'}^{-1}(s)$, i.e.\
$s_\text{pole}=-\i\varepsilon_k$, and we are allowed to expand the
nonsingular second factor as a power series of $\lambda$, yielding
\begin{equation}
G_{\bm{k}\bm{k}'}(t)=\delta^3(\bm{k}-\bm{k}')\e^{-\i\varepsilon_kt}+O(\lambda^2).
\label{eqn:Glowest}
\end{equation}

\section{Two-Particle Distribution in a Noncollinear Arrangement}\label{AppST}
Here we briefly sketch the derivation of (\ref{OffCorrelation}).
Remembering
\begin{subequations}
\label{B1}
\begin{gather}
\bar{\bm{r}}_1 = (\bar{r}_1\sin\Theta_\text{d} \cos\Phi, \bar{r}_1\sin\Theta_\text{d} \sin\Phi, \bar{r}_1\cos\Theta_\text{d})
,\\
\bar{\bm{r}}_2 = (-\bar{r}_2\sin\Theta_\text{d} \cos\Phi, -\bar{r}_2\sin\Theta_\text{d} \sin\Phi, \bar{r}_2\cos\Theta_\text{d}),
\end{gather}
\end{subequations}
the functions
$\mathcal{Z}_{k_1k_2}(\bar{\bm{r}})$ and
$\mathcal{A}_{\bm{k}_1\bm{k}_2}(\bar{\bm{r}})$
read
\begin{subequations}
\label{B3}
\begin{gather}
\mathcal{Z}_{k_1k_2}(\bar{\bm{r}})
=\theta(k_{1z})f(k_1\hat{\bar{\bm{r}}} )
\theta(k_{2z})f(k_2\hat{\bar{\bm{r}}} )
\e^{-(k_1-k_2)^2\hat{\bar{\bm{r}}}\cdot\mathcal{D}^2\hat{\bar{\bm{r}}}/2} ,
\\
\mathcal{A}_{\bm{k}_1\bm{k}_2}(\bar{\bm{r}})
=\frac{m^2}{(2\pi)^5}
\e^{-(\Delta\bm{k}_{1\hat{\bar{\bm{r}}}}\cdot\mathcal{W}^2\Delta\bm{k}_{1\hat{\bar{\bm{r}}}}+\Delta\bm{k}_{2\hat{\bar{\bm{r}}}}
\cdot\mathcal{W}^2\Delta\bm{k}_{2\hat{\bar{\bm{r}}}})/2} .
\end{gather}
\end{subequations}
By substituting (\ref{B3}) into (\ref{eqn:rho1}) and (\ref{eqn:I})
and introducing spherical coordinates, we get
\begin{equation}
\bar{\rho}_\text{st}(\bar{\bm{r}})
=\lambda^2\frac{2m^2}{(2\pi)^5\bar{r}^2}
\int_0^\infty\d k\,k^2N(\omega_k) f^2(k\hat{\bar{\bm{r}}})
\int_0^{\pi/2}\d\theta\sin\theta
\int_0^{2\pi}\d\phi\,
\e^{-\Delta\bm{k}_{\hat{\bar{\bm{r}}}}\cdot\mathcal{W}^2 \Delta\bm{k}_{\hat{\bar{\bm{r}}}}},
\end{equation}
\begin{multline}
\bar{\mathcal{I}}_\text{st}(\bar{\bm{r}}_1,\bar{\bm{r}}_2)
=\lambda^4\frac{2m^4}{(2\pi)^{10}\bar{r}_1^2\bar{r}_2^2}
\int_0^\infty\d k_1\,k_1^2 \int_0^\infty  \d k_2\,k_2^2 N(\omega_{k_1}) N(\omega_{k_2})
f(k_1\hat{\bar{\bm{r}}}_1)f(k_2\hat{\bar{\bm{r}}}_1)
f(k_1\hat{\bar{\bm{r}}}_2)f(k_2\hat{\bar{\bm{r}}}_2)\\
{}\times
\e^{-\i(k_1-k_2)(\bar{r}_1-\bar{r}_2)}
\e^{-(k_1-k_2)^2(\hat{\bar{\bm{r}}}_1\cdot\mathcal{D}^2\hat{\bar{\bm{r}}}_1+
\hat{\bar{\bm{r}}}_2\cdot\mathcal{D}^2\hat{\bar{\bm{r}}}_2)/2}
J(k_1;\bar{\bm{r}}_1,\bar{\bm{r}}_2) J(k_2;\bar{\bm{r}}_1,\bar{\bm{r}}_2),
\end{multline}
where
\begin{equation}
J(k;\bar{\bm{r}}_1,\bar{\bm{r}}_2)
=\int_0^{\pi/2}
\d\theta\sin\theta \int_0^{2\pi}\d\phi\,
\e^{-(\Delta\bm{k}_{\hat{\bar{\bm{r}}}_1}\cdot\mathcal{W}^2\Delta\bm{k}_{\hat{\bar{\bm{r}}}_1}+\Delta\bm{k}_{\hat{\bar{\bm{r}}}_2}\cdot\mathcal{W}^2 \Delta\bm{k}_{\hat{\bar{\bm{r}}}_2})/2}.
\end{equation}
Note that, when $\hat{\bar{\bm{r}}}_1$ and $\hat{\bar{\bm{r}}}_2$ satisfy (\ref{B1}), one has
\begin{equation}
f(k\hat{\bar{\bm{r}}}_1)=f(k\hat{\bar{\bm{r}}}_2)
=\frac{1}{ [(2\pi)^3(\delta k_z)^2(\delta k_\perp)^4]^{1/4}}
\exp\!\left(-\frac{k^2\sin^2\!\Theta_\text{d}}{ 4(\delta k_\perp)^2}-\frac{(k\cos\Theta_\text{d}-k_0)^2 }{ 4(\delta k_z)^2}\right)
\equiv f(k,\Theta_\text{d}) ,
\end{equation}
and $\hat{\bar{\bm{r}}}_1\cdot\mathcal{D}^2\hat{\bar{\bm{r}}}_1
=\hat{\bar{\bm{r}}}_2\cdot\mathcal{D}^2\hat{\bar{\bm{r}}}_2=a^2\sin^2\!\Theta_\text{d}+d^2\cos^2\!\Theta_\text{d}$.

\subsection{First-Order Correlation}
Let us first consider the case where $\bm{r}=\bm{r}_1$. Since
\begin{equation}
\Delta\bm{k}_{\hat{\bar{\bm{r}}}}\cdot\mathcal{W}^2 \Delta\bm{k}_{\hat{\bar{\bm{r}}}}
=
w^2k^2(\sin^2\!\Theta_\text{d}+\sin^2\!\theta)
+w_z^2k^2(\cos\Theta_\text{d}-\cos\theta)^2
-2w^2k^2\sin\Theta_\text{d}\sin\theta\cos(\Phi-\phi),
\end{equation}
one has
\begin{equation}
\int_0^{\pi/2}
\d\theta\sin\theta
\int_0^{2\pi}\d\phi\,
\e^{-\Delta\bm{k}_{\hat{\bar{\bm{r}}}}\cdot\mathcal{W}^2\Delta\bm{k}_{\hat{\bar{\bm{r}}}}}
=
\e^{-p\sin^2\!\Theta_\text{d}}
\int_0^{\pi/ 2}\d\theta\sin\theta
\, \e^{-p\sin^2\!\theta-q(\cos\Theta_\text{d}-\cos\theta)^2}
\int_0^{2\pi}\d\phi\, \e^{2p \sin\Theta_\text{d}\sin\theta\cos(\Phi-\phi)} ,
\end{equation}
where $p=w^2k^2$ and $q=w_z^2k^2$.
   The integral over $\phi$ yields a modified Bessel function of the first kind:
   \begin{align}
   \int_0^{2\pi}\d\phi\,\e^{2p\sin\Theta_\text{d}\sin\theta\cos(\Phi-\phi)}
   =2\pi I_0(2p\sin\Theta_\text{d}\sin\theta)
   = 2\pi\sum_{m=0}^\infty \frac{(p\sin\Theta_\text{d}\sin\theta)^{2m}}{ (m!)^2}.
   \end{align}
Therefore, with the aid of the trick used in
(\ref{eq:Iint}), we have
\begin{align}
&\int_0^{\pi/ 2}\d\theta\sin\theta
\int_0^{2\pi}\d\phi\,
\e^{-\Delta\bm{k}_{\hat{\bar{\bm{r}}}}\cdot\mathcal{W}^2 \Delta\bm{k}_{\hat{\bar{\bm{r}}}}}
\nonumber \\
&\qquad\qquad
=
2\pi\e^{-p\sin^2\!\Theta_\text{d}}
\sum_{m=0}^\infty \frac{(p^2\sin^2\!\Theta_\text{d})^m}{(m!)^2}
\int_0^{\pi/ 2} \d\theta \sin^{2m+1}\!\theta\,
\e^{-p\sin^2\!\theta-q(\cos\Theta_\text{d}-\cos\theta)^2}
\nonumber\displaybreak[0]\\
&\qquad\qquad
=
2\pi \e^{-p\sin^2\!\Theta_\text{d}}
\sum_{m=0}^\infty \frac{(p^2\sin^2\!\Theta_\text{d})^m}{(m!)^2}
\left(-\frac{\partial}{\partial p}\right)^m
\int_0^{\pi/ 2} \d\theta \sin\theta\,
\e^{-p\sin^2\!\theta-q(\cos\Theta_\text{d}-\cos\theta)^2}
.
\label{AngInt}
\end{align}
Hereafter, the integral over $\theta$ is evaluated when $p\gg q$ and
$p\gg 1$. By changing the integration variable, we have
\begin{equation}
\int_0^{\pi/ 2} \d\theta \sin\theta
\,\e^{-p\sin^2\!\theta-q(\cos\Theta_\text{d}-\cos\theta)^2}
=
\frac{\e^{-p\left(1+\frac{q\cos^2\!\Theta_\text{d}}{p-q}\right)}}{ \sqrt{p-q}}\,\Biggl(
\int_{\frac{q\cos\Theta_\text{d}}{\sqrt{p-q}}}^1 \d x \,\e^{x^2}
+
\int_1^{\sqrt{p-q}+\frac{q\cos\Theta_\text{d}}{\sqrt{p-q}}} \d x \,\e^{x^2}
\Biggr).
\end{equation}
The first term is of order of $\e^{-p}/\sqrt{p}$. Integrating by
parts, the second term becomes
\begin{multline}
\frac{\e^{-p\left(1+\frac{q\cos^2\!\Theta_\text{d}}{p-q}\right)}}{ \sqrt{p-q}}
\int_1^{\sqrt{p-q}+\frac{q\cos\Theta_\text{d}}{\sqrt{p-q}}} \d x \,\e^{x^2}
=\frac{ \e^{-q(\cos\Theta_\text{d}-1)^2}}{ 2(p-q+q\cos\Theta_\text{d})}\left(
1+\frac{1}{2\left(\sqrt{p-q}+\frac{q\cos\Theta_\text{d}}{\sqrt{p-q}}\right)^2}
\right)
\\
-\frac{3\e}{4}
\frac{\e^{-p\left(1+\frac{q\cos^2\!\Theta_\text{d}}{p-q}\right)}}{ \sqrt{p-q}}
{}+
\frac{\e^{-p\left(1+\frac{q\cos^2\!\Theta_\text{d}}{p-q}\right)}}{ \sqrt{p-q}}
\int_1^{\sqrt{p-q}+\frac{q\cos\Theta_\text{d}}{\sqrt{p-q}}} \d x \,\frac{3\e^{x^2} }{4x^4}
,
\end{multline}
where the second and third terms are, respectively, $O(1/p^{3/2})$ and $O(\e^{-p}/\sqrt{p})$,
and the last term can be shown to be O$(p^{-5/4})$.
Thus,
one has
\begin{align}
\int_0^{\pi/ 2} \d\theta \sin\theta\,
\e^{-p\sin^2\!\theta-q(\cos\Theta_\text{d}-\cos\theta)^2}
&=
\frac{ \e^{-q(\cos\Theta_\text{d}-1)^2}}{ 2(p-q+q\cos\Theta_\text{d})}
+O(p^{-{5/4}})
\nonumber\\
&=
\frac{ \e^{-q(\cos\Theta_\text{d}-1)^2}}{2p}
+O(p^{-{5/ 4}}). \label{AzInt}
\end{align}

By substituting (\ref{AzInt}) into (\ref{AngInt}) and retaining the leading order terms
in $1/p$, we have
\begin{align}
\int_0^{\pi/ 2}\d\theta\sin\theta
\int_0^{2\pi}\d\phi\,
\e^{-\Delta\bm{k}_{\hat{\bar{\bm{r}}}}\cdot\mathcal{W}^2 \Delta\bm{k}_{\hat{\bar{\bm{r}}}}}
&=2
\pi \e^{-p\sin^2\!\Theta_\text{d}}
\sum_{m=0}^\infty \frac{(p^2\sin^2\!\Theta_\text{d})^m}{(m!)^2}
\left(-\frac{\partial}{ \partial p}\right)^m
\frac{\e^{-q(\cos\Theta_\text{d}-1)^2} }{ 2p}
\nonumber \\
&
=
\pi \e^{-p\sin^2\!\Theta_\text{d}}
\e^{-q(\cos\Theta_\text{d}-1)^2}
\sum_{m=0}^\infty \frac{(p^2\sin^2\!\Theta_\text{d})^m}{(m!)^2}
\frac{m!}{ p^{m+1}}
\nonumber\\
&
=
\frac{\pi}{ p} \e^{-p\sin^2\!\Theta_\text{d}-q(1-\cos\Theta_\text{d})^2}
\e^{p\sin^2\!\Theta_\text{d}}
\nonumber \\
&=
\frac{\pi}{w^2k^2}\e^{-w_z^2k^2(1-\cos\Theta_\text{d})^2}.
\end{align}
Then, the one-particle correlation is
\begin{equation}
\bar{\rho}_\text{st}(\bar{\bm{r}})
=\lambda^2
\frac{m^2}{ (2\pi)^4w^2\bar{r}^2}
\int_0^\infty \d k\, N(\omega_k) f^2(k,\Theta_\text{d})
\e^{-w_z^2k^2(1-\cos\Theta_\text{d})^2} .
\label{B16}
\end{equation}
When $\bar{\bm{r}}=\bar{\bm{r}}_2$, we obtain the same result. This
is the denominator of Eqs.\ (\ref{eqn:CorrInterfere}) and
(\ref{OffCorrelation}).

\subsection{Second-Order Correlation}
Because (\ref{B1}) leads to
\begin{equation}
\frac{1}{ 2}
(\Delta\bm{k}_{\hat{\bar{\bm{r}}}_1}\cdot\mathcal{W}^2\Delta\bm{k}_{\hat{\bar{\bm{r}}}_1}
+
\Delta\bm{k}_{\hat{\bar{\bm{r}}}_2}\cdot\mathcal{W}^2\Delta\bm{k}_{\hat{\bar{\bm{r}}}_2})
=p(\sin^2\!\Theta_\text{d}+\sin^2\!\theta)+
q(\cos\Theta_\text{d}-\cos\theta)^2 ,
\end{equation}
the auxiliary function $J(k;\bar{\bm{r}}_1,\bar{\bm{r}}_2)$ is evaluated as
\begin{align}
J(k;\bar{\bm{r}}_1,\bar{\bm{r}}_2)
&= 2\pi \int_0^{\pi/2} \d\theta\sin\theta\, \e^{-p(\sin^2\!\Theta_\text{d}+\sin^2\!\theta)-q(\cos\Theta_\text{d}-\cos\theta)^2}
\nonumber\\
&= \frac{\pi}{w^2k^2}\e^{-w^2k^2\sin^2\!\Theta_\text{d}}
\e^{-w_z^2k^2(1-\cos\Theta_\text{d})^2},
\end{align}
where we have used (\ref{AzInt}) in the second equality.
Thus, in terms of $f(k_j,\Theta_\text{d})$ ($j=1,2$),
\begin{multline}
\bar{\mathcal{I}}_\text{st}(\bar{\bm{r}}_1,\bar{\bm{r}}_2)
=\lambda^4\frac{m^4}{2(2\pi)^8w^4\bar{r}_1^2\bar{r}_2^2 }
\int_0^\infty \d k_1 \int_0^\infty \d k_2\, N(\omega_{k_1}) N(\omega_{k_2})
f^2(k_1,\Theta_\text{d}) f^2(k_2,\Theta_\text{d})
\e^{-\i(k_1-k_2)(\bar{r}_1-\bar{r}_2)}
\\
{}\times \e^{-(k_1-k_2)^2(a^2\sin^2\!\Theta_\text{d}+d^2\cos^2\!\Theta_\text{d})}
\e^{-(k_1^2+k_2^2)[w^2\sin^2\!\Theta_\text{d}+w_z^2(1-\cos\Theta_\text{d})^2]}.
\label{B19}
\end{multline}
This is the numerator of Eqs.\ (\ref{eqn:CorrInterfere}) and
(\ref{OffCorrelation}).

\subsection{Well-Monochromatized Case}
If the beam of particles is well-monochromatized and the distribution $N(\omega_k)$ is a slowly varying function there, we have
\begin{align}
\bar{\rho}_\text{st}(\bar{\bm{r}})
&=
\lambda^2\frac{m^2}{ (2\pi)^4w^2\bar{r}^2}
N(\omega_{k_0})
\int_{-\infty}^\infty \d k\, f^2(k,\Theta_\text{d})
\e^{-w_z^2k^2(1-\cos\Theta_\text{d})^2}
\nonumber \\
&=\lambda^2\frac{m^2}{ (2\pi)^5w^2\bar{r}^2(\delta k_\perp)\sqrt{D_1(\Theta_\text{d})}}
  N(\omega_{k_0})
  \exp\!\left[
  -\frac{k_0^2}{ 2 (\delta k_z)^2}
  \left(
  1-\frac{(\delta k_\perp)^2 \cos^2\!\Theta_\text{d} }{ D_1(\Theta_\text{d})}
  \right)
  \right],
\end{align}
where the Gaussian $k$-integration has been carried out with the aid of
\begin{align}
-\frac{k^2\sin^2\!\Theta_\text{d}}{ 2(\delta k_\perp)^2}
-\frac{k^2\cos^2\!\Theta_\text{d}-2k_0k\cos\Theta_\text{d}+k_0^2 }{ 2(\delta k_z)^2}
&-w_z^2k^2(1-\cos\Theta_\text{d})^2
\nonumber\\
&=
-
\frac{D_1(\Theta_\text{d})}{ 2(\delta k_z)^2(\delta k_\perp)^2}
k^2
+\frac{k_0\cos\Theta_\text{d} }{ (\delta k_z)^2} k
-\frac{k_0^2}{ 2(\delta k_z)^2}.
\end{align}
On the other hand, we have
\begin{multline}
\bar{\mathcal{ I}}_\text{st}(\bar{\bm{r}}_1,\bar{\bm{r}}_2)
=\lambda^4\frac{m^4}{2(2\pi)^8w^4\bar{r}_1^2\bar{r}_2^2}
N^2(\omega_{k_0})
\int_{-\infty}^\infty \d k_1 \int_{-\infty}^\infty \d k_2\,
f^2(k_1,\Theta_\text{d})f^2(k_2,\Theta_\text{d})
\e^{-\i(k_1-k_2)(\bar{r}_1-\bar{r}_2)}
\\
{}\times
\e^{-(k_1-k_2)^2(a^2\sin^2\!\Theta_\text{d}+d^2\cos^2\!\Theta_\text{d})}
\e^{-(k_1^2+k_2^2)[w^2\sin^2\!\Theta_\text{d}+w_z^2(1-\cos\Theta_\text{d})^2]}.
\end{multline}
In terms of $K=(k_1+k_2)/2$ and $k=k_1-k_2$, one has
\begin{multline}
-\sum_{j=1}^2 \left(
\frac{k_j^2\sin^2\!\Theta_\text{d}}{ 2(\delta k_\perp)^2}
+\frac{k_j^2\cos^2\!\Theta_\text{d}-2k_0k_j\cos\Theta_\text{d}+k_0^2 }{ 2(\delta k_z)^2}
\right)
-\i(k_1-k_2)(\bar{r}_1-\bar{r}_2)
\\
{}-(k_1-k_2)^2(a^2\sin^2\!\Theta_\text{d}+d^2\cos^2\!\Theta_\text{d})
-(k_1^2+k_2^2)[w^2\sin^2\!\Theta_\text{d}+w_z^2(1-\cos\Theta_\text{d})^2]
\\
=
-\frac{D_2(\Theta_\text{d})}{(\delta k_z)^2(\delta k_\perp)^2}
K^2 + \frac{2k_0\cos\Theta_\text{d}}{ (\delta k_z)^2}K
-\frac{k_0^2}{ (\delta k_z)^2}
-\frac{D_3(\Theta_\text{d})}{ 4(\delta k_z)^2(\delta k_\perp)^2}
k^2
-\i k(\bar{r}_1-\bar{r}_2),
\end{multline}
and, thus,
\begin{multline}
\bar{\mathcal{ I}}_\text{st}(\bar{\bm{r}}_1,\bar{\bm{r}}_2)
=\lambda^4\frac{m^4 }{ 2(2\pi)^{10}w^4\bar{r}_1^2 \bar{r}_2^2(\delta k_\perp)^2
\sqrt{D_2(\Theta_\text{d})D_3(\Theta_\text{d})}}
N^2(\omega_{k_0})
\\
{}\times
\exp\!\left[
-\frac{k_0^2}{ (\delta k_z)^2}\left(
1-\frac{(\delta k_\perp)^2\cos^2\!\Theta_\text{d}}{ D_2(\Theta_\text{d})}
\right)
-\frac{(\delta k_z)^2(\delta k_\perp)^2(\bar{r}_1-\bar{r}_2)^2}{ D_3(\Theta_\text{d})}
\right].
\end{multline}
By plugging (\ref{B16}) and (\ref{B19}) into (\ref{eqn:CorrInterfere}), we
obtain the normalized two-particle distribution function
(\ref{OffCorrelation}).

\end{document}